\documentclass[ALICE,manyauthors]{cernphprep}
\usepackage[comma,square,numbers,sort&compress]{natbib}
\usepackage{lineno}
\usepackage{xspace}
\usepackage{xcolor,soul}
\usepackage{hyperref}
\usepackage{xparse}
\usepackage{xstring}
\usepackage{euscript}
\usepackage[utf8]{inputenc}
\usepackage[english]{babel}
\usepackage{graphicx}
\usepackage{blindtext}
\usepackage{float}
\usepackage{adjustbox}
\usepackage{enumitem}

\begin{document}

\newcommand{\pp}           {pp\xspace}
\newcommand{\ppbar}        {\mbox{$\mathrm {p\overline{p}}$}\xspace}
\newcommand{\XeXe}         {\mbox{Xe--Xe}\xspace}
\newcommand{\PbPb}         {\mbox{Pb--Pb}\xspace}
\newcommand{\pA}           {\mbox{pA}\xspace}
\newcommand{\pPb}          {\mbox{p--Pb}\xspace}
\newcommand{\AuAu}         {\mbox{Au--Au}\xspace}
\newcommand{\dAu}          {\mbox{d--Au}\xspace}

\newcommand{\s}            {\ensuremath{\sqrt{s}}\xspace}
\newcommand{\snn}          {\ensuremath{\sqrt{s_{\mathrm{NN}}}}\xspace}
\newcommand{\pt}           {\ensuremath{p_{\rm T}}\xspace}
\newcommand{\meanpt}       {$\langle p_{\mathrm{T}}\rangle$\xspace}
\newcommand{\ycms}         {\ensuremath{y_{\rm CMS}}\xspace}
\newcommand{\ylab}         {\ensuremath{y_{\rm lab}}\xspace}
\newcommand{\etarange}[1]  {\mbox{$\left | \eta \right |~<~#1$}}
\newcommand{\yrange}[1]    {\mbox{$\left | y \right |~<~#1$}}
\newcommand{\dndy}         {\ensuremath{\mathrm{d}N/\mathrm{d}y}\xspace}
\newcommand{\dnchdy}       {\ensuremath{\mathrm{d}N_\mathrm{ch}/\mathrm{d}y}\xspace}
\newcommand{\dndeta}       {\ensuremath{\mathrm{d}N/\mathrm{d}\eta}\xspace}
\newcommand{\dnchdeta}     {\ensuremath{\mathrm{d}N_\mathrm{ch}/\mathrm{d}\eta}\xspace}
\newcommand{\avdndeta}     {\ensuremath{\langle\dndeta\rangle}\xspace}
\newcommand{\dNdy}         {\ensuremath{\mathrm{d}N/\mathrm{d}y}\xspace}
\newcommand{\Npart}        {\ensuremath{N_\mathrm{part}}\xspace}
\newcommand{\meanNpart}    {$\langle N_\mathrm{part}\rangle$\xspace}
\newcommand{\Ncoll}        {\ensuremath{N_\mathrm{coll}}\xspace}
\newcommand{\meanNcoll}    {$\langle N_\mathrm{coll}\rangle$\xspace}
\newcommand{\dEdx}         {\ensuremath{\textrm{d}E/\textrm{d}x}\xspace}
\newcommand{\RpPb}         {\ensuremath{R_{\rm pPb}}\xspace}
\newcommand{\RAA}          {\ensuremath{R_{\rm AA}}\xspace}
\newcommand{\Tc}           {\ensuremath{T_{\rm c}}\xspace}
\newcommand{\Tch}          {\ensuremath{T_{\rm ch}}\xspace}
\newcommand{\Tfo}          {\ensuremath{T_{\rm fo}}\xspace}
\newcommand{\Tkin}         {\ensuremath{T_{\rm kin}}\xspace}
\newcommand{\muB}          {\ensuremath{\mu_{\rm B}}\xspace}

\newcommand{\nineH}        {$\sqrt{s}~=~0.9$~Te\kern-.1emV\xspace}
\newcommand{\seven}        {$\sqrt{s}~=~7$~Te\kern-.1emV\xspace}
\newcommand{\twoH}         {$\sqrt{s}~=~0.2$~Te\kern-.1emV\xspace}
\newcommand{\twosevensix}  {$\sqrt{s}~=~2.76$~Te\kern-.1emV\xspace}
\newcommand{\five}         {$\sqrt{s}~=~5.02$~Te\kern-.1emV\xspace}
\newcommand{\twosevensixnn}{$\sqrt{s_{\mathrm{NN}}}~=~2.76$~Te\kern-.1emV\xspace}
\newcommand{\fivenn}       {$\sqrt{s_{\mathrm{NN}}}~=~5.02$~Te\kern-.1emV\xspace}
\newcommand{\fivennt}       {$\sqrt{s_{\mathrm{NN}}}~=~5.36$~Te\kern-.1emV\xspace}
\newcommand{\LT}           {L{\'e}vy-Tsallis\xspace}
\newcommand{\GeVc}         {Ge\kern-.1emV/$c$\xspace}
\newcommand{\MeVc}         {Me\kern-.1emV/$c$\xspace}
\newcommand{\TeV}          {Te\kern-.1emV\xspace}
\newcommand{\GeV}          {Ge\kern-.1emV\xspace}
\newcommand{\MeV}          {Me\kern-.1emV\xspace}
\newcommand{\GeVmass}      {Ge\kern-.2emV/$c^2$\xspace}
\newcommand{\MeVmass}      {Me\kern-.2emV/$c^2$\xspace}
\newcommand{\lumi}         {\ensuremath{\mathcal{L}}\xspace}

\newcommand{\ITS}          {\rm{ITS}\xspace}
\newcommand{\TOF}          {\rm{TOF}\xspace}
\newcommand{\ZDC}          {\rm{ZDC}\xspace}
\newcommand{\ZDCs}         {\rm{ZDCs}\xspace}
\newcommand{\ZNA}          {\rm{ZNA}\xspace}
\newcommand{\ZNC}          {\rm{ZNC}\xspace}
\newcommand{\SPD}          {\rm{SPD}\xspace}
\newcommand{\SDD}          {\rm{SDD}\xspace}
\newcommand{\SSD}          {\rm{SSD}\xspace}
\newcommand{\TPC}          {\rm{TPC}\xspace}
\newcommand{\TRD}          {\rm{TRD}\xspace}
\newcommand{\VZERO}        {\rm{V0}\xspace}
\newcommand{\VZEROA}       {\rm{V0A}\xspace}
\newcommand{\VZEROC}       {\rm{V0C}\xspace}
\newcommand{\Vdecay} 	   {\ensuremath{V^{0}}\xspace}

\newcommand{\ee}           {\ensuremath{e^{+}e^{-}}} 
\newcommand{\pip}          {\ensuremath{\pi^{+}}\xspace}
\newcommand{\pim}          {\ensuremath{\pi^{-}}\xspace}
\newcommand{\kap}          {\ensuremath{\rm{K}^{+}}\xspace}
\newcommand{\kam}          {\ensuremath{\rm{K}^{-}}\xspace}
\newcommand{\pbar}         {\ensuremath{\rm\overline{p}}\xspace}
\newcommand{\kzero}        {\ensuremath{{\rm K}^{0}_{\rm{S}}}\xspace}
\newcommand{\lmb}          {\ensuremath{\Lambda}\xspace}
\newcommand{\almb}         {\ensuremath{\overline{\Lambda}}\xspace}
\newcommand{\Om}           {\ensuremath{\Omega^-}\xspace}
\newcommand{\Mo}           {\ensuremath{\overline{\Omega}^+}\xspace}
\newcommand{\X}            {\ensuremath{\Xi^-}\xspace}
\newcommand{\Ix}           {\ensuremath{\overline{\Xi}^+}\xspace}
\newcommand{\Xis}          {\ensuremath{\Xi^{\pm}}\xspace}
\newcommand{\Oms}          {\ensuremath{\Omega^{\pm}}\xspace}
\newcommand{\degree}       {\ensuremath{^{\rm o}}\xspace}

\begin{titlepage}

\PHyear{2022}       
\PHnumber{XXX}      
\PHdate{1st December}  


\title{Hot and Cold QCD White Paper from ALICE-USA: \\ Input for the 2023 U.S. Long Range Plan for Nuclear Science}
\ShortTitle{Hot \& Cold QCD plans for 2023 Long Range Plan for Nuclear Science}   

\Collaboration{The ALICE-USA Collaboration\thanks{See Appendix~\ref{app:collab} for the list of collaboration members}}
\ShortAuthor{ALICE-USA Collaboration}

\end{titlepage}


\setcounter{page}{2} 


\section{Executive Summary}

The ALICE experiment was built to study many-body Quantum Chromo-Dynamics (QCD) at high temperature and effectively zero baryon density, using relativistic heavy-ion collisions at the Large Hadron Collider (LHC). These collisions form the Quark Gluon Plasma (QGP), a state of matter where quarks and gluons are no longer confined inside hadrons. The ALICE program centers around the key questions related to QGP phenomena. These include the macroscopic and microscopic properties of the QGP, and the details of the QGP phase transition to hadrons, that is believed to have taken place in the early Universe. At the same time, ALICE's versatile setup allows for the study of \pp collisions, \pPb collisions, and ultra-peripheral collisions. The associated studies can provide some of the most stringent tests of QCD and Beyond Standard Model searches. They serve as deep probes of the properties of cold nuclear matter, and allow for investigations of stellar and interstellar phenomena. The ALICE-USA collaboration consists of 13 institutes, representing about 6\% of the total ALICE authorship. It has, and continues to play, an essential role in all areas of ALICE leadership: physics, instrumentation, and management. ALICE-USA has been involved in $\sim$25\% of ALICE's $\sim$400 ALICE papers since 2009, including many of ALICE's most cited and high profile publications. We, the ALICE-USA collaboration, will provide two recommendations for the U.S. 2023 Long Range Plan for Nuclear Science period, and beyond. These recommendations are essential for maintaining the continued success and development of the U.S. Nuclear Science community for both Hot and Cold QCD. The recommendations will be accompanied by details of the vital ALICE-USA scientific priorities. We will also describe the broader impact and support needed for U.S. involvement in ALICE.

\paragraph{Recommendation 1: Continue and broaden the contribution from U.S. institutes for the ALICE program in Runs 3\&4.} 

ALICE has just completed a number of major upgrades. ALICE-USA has made vital contributions to the new Inner Tracking System (ITS) and Time Projection Chamber (TPC) readout. ALICE-USA will now utilize these upgrades for a comprehensive physics program. This `ALICE 2' phase also provides a unique opportunity for Hot and Cold QCD studies between the expected times when RHIC discontinues taking data in 2025, and when the EIC begins taking data around 2035. The new detector setup began taking data in Run 3 (2022-2025), and will continue to operate in Run 4 (2029-2032). During these periods, ALICE will collect data from collisions of: \PbPb at $\sqrt{s_{\mathrm{NN}}}=5.36$~TeV, \pp at $\sqrt{s}=13.6$~TeV, \pp at the same energy as \PbPb, \pPb at $\sqrt{s_{\mathrm{NN}}}=8$~TeV, p--O at $\sqrt{s_{\mathrm{NN}}}=9.9$~TeV, and O--O at $\sqrt{s_{\mathrm{NN}}}=7$~TeV. The new data set of Pb--Pb collisions ($\sim$13 nb$^{-1}$) represents an increase of two orders of magnitude over Runs 1\&2. These increased capabilities in Runs 3\&4 will allow for:
\begin{itemize}
 
\item The elucidation of the microscopic parton dynamics underlying QGP properties using hard processes of lower cross section, such as heavy-flavor jets, over a large range of transverse momenta;

\item The characterization of the macroscopic QGP properties with extended precision, including the exploration of unknown dynamical QGP transport parameters at LHC energies, such as the the baryon diffusion coefficient;

\item Deeper studies of the hadronization of heavy-flavor baryons and mesons produced in high temperature QCD matter;

\item Multiple observables with an increased precision to contribute to the development of a unified picture of particle production and QCD dynamics from small (pp) to large (nucleus–nucleus);

\item Unique explorations of parton densities in protons and nuclei in a broad ($x$, $Q^2$) kinematic range, reaching to $x\sim 10^{-6}$. 

\end{itemize}

In relation to the last point, ALICE-USA is one of the key proponents of the Forward Calorimeter (FoCal). This a new detector that will collect data in Run 4. The development, installation, and operation of the FoCal will occur during the LRP duration, and prior to the start of the EIC data taking period. The FoCal is designed to provide unique capabilities to probe the structure of nucleons and nuclei in an unexplored regime of $x$ and $Q^2$. This kinematic region, of $x\sim 10^{-6}$, will probe the regime where gluon saturated matter is expected to be dominant. Such a small-$x$ region is not accessible at the EIC. Global fits of this lower $x$ data, with new high precision EIC data, will make an unprecedented step to understanding the full evolution of Parton Distributions Functions (PDFs) in protons and nuclei. The R\&D and design phase of the EIC detectors have well defined synergies with the FoCal instrumentation work. We also stress the additional benefits of U.S. involvement in ALICE's instrumentation R\&D for the ITS3 sensors that will be part of the Run 4 setup. These have various applications for the EIC detector projects, as recognized in multiple forums. Support for both of these endeavors is of vital importance. 

Beyond the scientific and technological implications discussed, the active participation of U.S. institutions in the ALICE collaboration in the coming years would greatly benefit the training of the next generation of scientists. It is vital that PhD students, postdoctoral researchers, and early career faculty, have the opportunity to develop their data analysis, modeling, and interpretation expertise using the largest and most complex datasets of QCD phenomena ever garnered. These younger members of the community will also have the opportunity to participate in the full cycle of detector operation and development in the 2020s. Such experience is essential for maintaining the U.S. as a leader in nuclear science. Therefore, continued support for ALICE research is crucial. We urge a modest expansion of the ALICE efforts to allow groups currently focused at RHIC to also participate prior to EIC operations.

\paragraph{Recommendation 2: Begin participation in ALICE-3 R\&D and construction.}

ALICE-USA fully supports the ALICE 3 detector, a next generation detector that is designed to operate in Runs 5\&6 (2035 and beyond). It is currently in the R\&D phase, and construction of the ALICE 3 detector is expected to begin 2028. It represents a unique direction to pursue Hot QCD studies at the highest temperatures possible into and beyond the 2030s. ALICE 3 will have the largest acceptance and most precise tracking ever achieved for a heavy-ion experiment. It is expected to record  $\sim$20 times more heavy-ion data compared to Runs 3\&4. The general purpose detector design of ALICE 3 will provide many physics opportunities, which are expected to lead to new discoveries. It will also provide a level of data quality required to transform our understanding of the QGP, in terms of first principles QCD calculations. The key topics unique to ALICE 3, and under the umbrella of current ALICE-USA interests, include:
\begin{itemize}
 
\item Jet hadrochemistry over an unprecedented kinematic range for transverse momentum and rapidity, to provide new ways to investigate QGP jet energy loss and the microscopic structure of the QGP;

\item Measurements of multi-heavy-flavor and exotic charm states to provide unique information on hadronization and hadronic interactions at high temperatures; 

\item Beyond Standard Model physics with searches for axion-like particles in ultra-peripheral collisions, which will leverage the high collision rates and large acceptance.

\end{itemize}

The design of ALICE 3 will also enable measurements of high-precision, multi-differential of electromagnetic radiation from the QGP to probe its early evolution and the restoration of chiral symmetry through the coupling of vector and axial-vector mesons. Measurements of net-quantum number fluctuations over a wide rapidity range will constrain the susceptibilities of the QGP, and test the realization of the cross-over phase transition expected at LHC energies. Finally, for the first time, measurements of the production of ultra-soft photons will be possible. These can quantitatively test the infrared limits of quantum field theories such as QED and QCD. In order to meet these physics goals, support for both the R\&D and construction efforts is critically required from the U.S. community. This needs to begin before the end of the current 2023 Long Range Plan for Nuclear Science. Based on its expertise, ALICE-USA is currently exploring contributions to the silicon and calorimeter detectors. These parallel and independent efforts will also benefit the EIC achieving its construction milestones.

\section{Overview of ALICE upgrades and U.S. impact}

ALICE is a large acceptance experiment, with world leading hadron identification capabilities. Unique to the LHC experiments, ALICE's primary focus, design, and proposed upgrades, enable extremely accurate characterizations of the QGP using a multitude of observables. It has extensive capabilities in exploring few-body hadronic and nuclear interactions. The first major upgrade for LHC Runs 3\&4 enables an increase of ALICE's recorded luminosities by two orders of magnitude via the introduction of a continuous readout system in the TPC. The second involves a greatly improved tracking performance using new inner trackers - ITS2 in Run 3, and ITS3 in Run 4. The ITS2 system has a reduction of a factor three in radiation lengths, and factor two improvement in pointing resolution compared to the ITS used in Runs 1\&2. Further factor of three improvements of the pointing resolution for ITS3 (compared to ITS2) will occur via the introduction of wafer-scale ultra-thin silicon strips, which are also being pursued by the EPIC detector at the EIC. Details of these upgrades and others can be found elsewhere~\cite{ALICE-PUBLIC-2019-001}. For Run 4, the Forward Calorimeter detector (FoCal) upgrade, with both electromagnetic and hadronic calorimeters, will provide unique capabilities for both Cold and Hot QCD studies at forward rapidities of $3.4 < \eta < 5.8$~\cite{ALICE-PUBLIC-2019-005}.

For Runs 5\&6, a completely new detector has been proposed, named ALICE 3~\cite{ALICE:2022wwr}. Designed by heavy-ion physicists for heavy-ion physics, it is a next generation detector with a main tracking system that covers a large pseudorapidity range ($-4<\eta<4$), and can reconstruct charged tracks down to $\pt \sim 100$ MeV/c. The light weight and small radiation length design will greatly reduce the background and improve tracking resolution for electromagnetic and heavy-flavor probes, compared to ALICE 2. The pointing resolution at midrapidity is projected to be about three times better than that of the ITS3, which is in part achieved by placing a highly novel vertex detector 5mm from the beamline. The tracking system is placed in a superconducting solenoid with a field of up to $B= 2$~T, to obtain a momentum resolution of 1--2\% over a broad pseudorapidity and momentum range. This tracking is complemented by multiple sub-detector systems for particle identification; two TOF detectors and a RICH detector. These systems have the ability to identify leptons and photons in the entire thermal emission range of $\pt \lesssim 3$ GeV/c, which is inaccessible for other LHC experiments. Charged hadron identification on the $3\sigma$ level is possible up to $\pt \sim 14$ GeV/c, and decay hadrons can be reconstructed much more efficiently and cleanly at low-\pt compared to ALICE 2. The fast readout systems will be able to record all of the expected heavy-ion luminosity provided by the LHC. The ALICE 3 program aims to collect an integrated luminosity of about 35~nb$^{-1}$ with Pb--Pb collisions and 18~fb$^{-1}$ with pp collisions at top LHC energy. The potential to further increase the luminosity for ion running in the LHC by using smaller ions, e.g.\,$^{84}$Kr or $^{128}$Xe, as well as further runs with small collision systems, is being explored.

\begin{figure}[ht]
\begin{center}
\includegraphics[width = 1\textwidth]{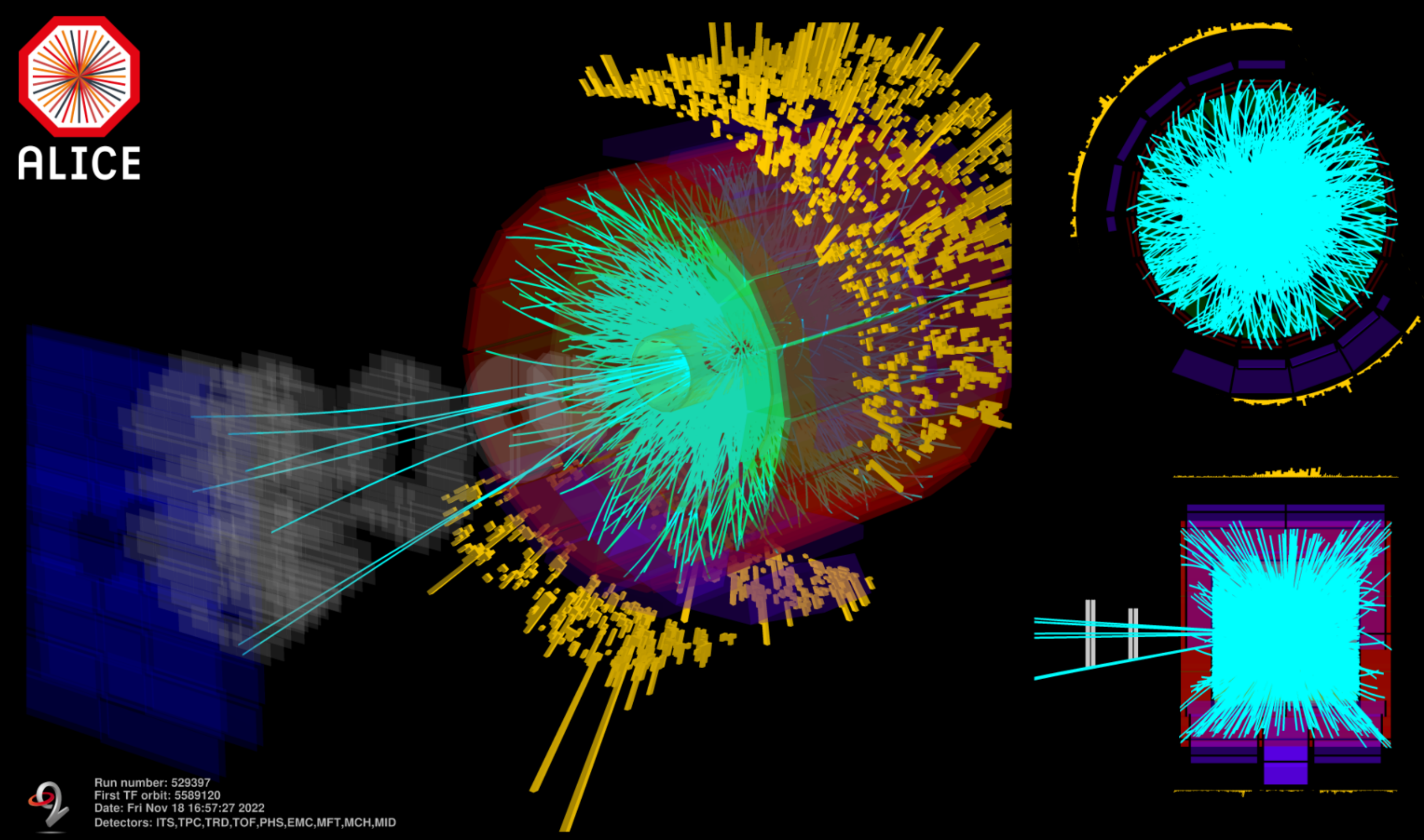}
\caption{Event display of one of the first LHC Run 3 \PbPb \fivennt collisions in November 18th 2022, using many detectors ALICE-USA has contributed toward the construction and upgrade.}
\end{center}
\label{Run3display}
\end{figure}

As of September 2022, the ALICE-USA groups consist of teams from two national laboratories: Lawrence Berkeley (LBNL) and Oak Ridge (ORNL), and nine universities: California (Berkeley), CalPoly, Chicago State, Creighton, Houston, Kansas, Ohio State, Tennessee (Knoxville), Texas (Austin), Wayne State, and Yale. While the majority of the program is supported by DOE, there are two NSF supported university teams. ALICE-USA has been responsible for two completed DOE funded projects - the EMCal/DCal in Runs 1\&2~\cite{ALICE:2803563}, and the most recently completed Barrel Tracking Upgrade (BTU) for Runs 3\&4 \footnote{See the BTU website at \url{https://sites.google.com/site/alicebtusite} for further details}. The BTU project involved critical U.S. contributions to both the TPC and ITS upgrades. A key example of the success of all these projects can be observed in Fig.~1, which shows one of the first ALICE Run 3 event-displays of \PbPb collisions using many detectors where ALICE-USA has made essential contributions. Thanks to the long and successful experience building and operating calorimeters, ALICE-USA is leading the FoCal project~\cite{ALICE:2022qhn}. It also oversees a major computing project providing the necessary U.S. contribution to ALICE's data processing infrastructure, under the umbrella of the Worldwide LHC Computing GRID (WLCG). 

In terms of the scientific output, the contributions from ALICE-USA span a wide range of physics topics within the collaboration, and have been an essential part of the ALICE physics program. Although ALICE-USA members are present at an about 6\% level in terms of authorship within the collaboration, they have provided leadership in terms of paper committee contributions well beyond the average contributions. ALICE-USA members were involved in 1 of 2 Nature articles published by ALICE~\cite{ALICE:2021aqk}, and 1 of 2 Nature Physics articles~\cite{ALICE:2016fzo}. ALICE-USA members were also involved in 5 of the top 7 most cited physics publications from ALICE~\cite{ALICE:2010suc,ALICE:2013snk,ALICE:2010mlf,ALICE:2011ab,ALICE:2016fzo}, and numerous other high-profile publications. These papers span both the hard and soft physics using heavy-ion and pp collisions. ALICE has published just over 400 papers since 2009, with the overall U.S. involvement on the paper committee level being about 25\%.

In addition to publications, members of the ALICE-USA have held numerous leadership roles across all seniority levels of the collaboration regarding physics output. These include: Deputy Physics Coordinator (two times), multiple Physics Working Group (PWG) Conveners, and Physics Analysis Group Coordinators. Of particular note, out of the 8 ALICE PWGs, currently ALICE-USA members co-convene 4 (50\%) of the those PWGs. Over the years members of ALICE-USA were copiously represented in the governance of ALICE -- currently with members of the collaboration holding functions as members of the Conference Committee and the Management Board, and as one of the two Deputy Spokespersons. Moreover, in the past ALICE-USA members served as Chair of the Collaboration Board and co-chair of the Editorial Board. Finally, for the $\sim100$ ALICE-USA members, roughly half are PhD students. At this rate, ALICE-USA institutions have graduated $\sim 50$ PhD students over a 5 year period. Many of our students have achieved leading roles in research positions in our field and beyond, as well as in industry and teaching jobs.
\newpage
\section{A brief summary of ALICE accomplishments from Runs 1\&2}

The observations ALICE has made during Runs 1\&2 (2009-2019) have profoundly changed the landscape of QCD at high temperatures and collision energies~\cite{ALICE:2022wpn}. They have been carried out in conjunction with other LHC experiments, and with the continuation of the RHIC program - where huge advances in luminosities and centre-of-mass energy coverage have been achieved. ALICE explorations of high temperature QCD have continued to reveal emergent behavior in many-body interactions at the highest possible temperatures in the laboratory. In heavy-ion collisions, ALICE has measured significant yield suppression for a wide range of hadrons and reconstructed jets in both inclusive and coincidence channels, showing that in-medium energy loss occurs at the partonic level, and quantifying its magnitude. ALICE measurements of heavy-flavor yield suppression provide insight into details of this process, via the QCD “dead-cone” effect. This manifests itself in a larger energy loss observed for charm compared to bottom quarks. It has also, for the time, provided direct evidence of the dead-cone effect for charm quarks in pp collisions, using state of the art jet re-clustering algorithms. ALICE jet substructure measurements indicate preferential suppression of wide-angle radiation in the jet shower within the QGP, which is sensitive to color coherence and the space-time structure of jets.

In the soft sector, ALICE measurements of identified hadron spectra and anisotropic flow demonstrate that a QGP formed at LHC energies undergoes the most rapid expansion ever observed for a many-body system in the laboratory. The hydrodynamic description of a huge variety of such data has been tested in heavy-ion collisions at the LHC, offering an environment far beyond the usual application in fluid dynamics. It has shown a QGP is strongly coupled at scales of the QGP's temperature, on the order of a few hundred MeV. ALICE measurements also imply thermalisation effects for charm quarks in a QGP, and when coupled with transport model calculations, the ALICE data demonstrate microscopically how equilibrium can occur on extremely small time scales. In addition, ALICE results have demonstrated that a QGP transitions into chemically equilibrated hadrons. ALICE has the most extensive set of measurements ever achieved regarding identified particle production, including for the first time a statistical description in the charm sector. ALICE investigations into the hadron-gas phase, via resonance and femtoscopic measurements, indicate this phase is prolonged, and that the decoupling of particles from the expanding hadron gas is likely to be a continuous process. 

Using RHIC and LHC Run 1\&2 data, comprehensive efforts at global fitting for precise determination of QGP properties and dynamics are currently underway, utilizing the powerful approach of Bayesian Inference for rigorous comparisons between theory and experiments. Such analyses are more advanced in the soft sector, and are in their infancy in the hard probes sector. ALICE has played a critical role in this endeavor by providing the experimental data, and developing theory-experiment synergies. Along with other RHIC and LHC experiments, ALICE has discovered QGP-like signatures in high multiplicity pp and p-Pb collisions, which probe the thresholds of QGP formation. Such findings have ignited a debate of whether pp and p-Pb collisions at LHC and RHIC energies could create small QGP droplets. 

ALICE has investigated few-body hadronic interactions on the soft and hard scales to an unprecedented precision. In pp collisions, it has demonstrated  charmed hadron fragmentation functions are not universal with respect to e+e- collisions, revealing unique hadronisation mechanisms that enhance charmed baryons compared to mesons. It also has provided world leading constraints regarding how rarely produced hadrons interact with stable nuclear matter, whose behavior have broad implications for understanding various features in the Universe, such as the equation of state of neutron stars and the interstellar composition of dark matter. Finally, ALICE has been a leader in the study of ultra-peripheral heavy-ion collisions since Run 1, exploring cold QCD matter at the highest possible photon-induced energies. ALICE has observed photoproduction of vector mesons on proton and ion targets, namely, the $J/\psi$, $\psi(2S)$, and $\rho$ mesons. $J/\psi$ photoproduction is moderately suppressed on Pb targets compared to free protons, in line with nuclear shadowing effects for gluon parton distribution functions at $x\approx 10^{-3}$. Coherent $\rho$ photoproduction measurements also demonstrate nuclear shadowing for both the Xe and Pb nuclei.


\section{Key physics questions addressed for LHC Runs 3\&4 and beyond from ALICE-USA}

Despite the many extraordinary scientific findings by the ALICE Collaboration, many open questions remain, and will be addressed using data to be collected in Runs 3\&4 at the LHC (2022-2032). This section details such questions, with a description of the leadership ALICE-USA groups intend to provide addressing them. Opportunities using the ALICE 3 detector in the Run 5\&6 period (2035 and beyond) will also be discussed.
\begin{enumerate}[leftmargin=*]
\item {\it What is the gluonic structure in protons \& nucleons at small-$x$?}

We propose to maximize the physics potential for small-$x$ physics in ALICE with the installation a high-granularity forward calorimeter~(FoCal) in the LHC Run~4~(2029--2032). The FoCal will explore the small-$x$ parton structure of nucleons and protons in a kinematic domain only accessible at the LHC~\cite{ALICE:2022qhn}. It is designed to provide unique capabilities to investigate Parton Distribution Functions~(PDFs) in the unexplored regime of  $x\sim10^{-6}$ and four momentum transfer $Q$ of a few \GeV. In this regime, it is expected that the hadronic structure requires non-linear dynamics due to the high gluon densities, leading to gluon saturation. Gluon saturation is a necessary consequence of the non-Abelian nature of QCD. Its discovery would represent a landmark in our understanding of the strong interaction. The primary objectives of the FoCal include: high-precision inclusive direct photon and jets measurements in \pp and \pPb collisions, photon-jet and jet-jet measurements in \pp and \pPb collisions, and measurements of vector mesons photoproduction in ultra-peripheral \pPb and \PbPb collisions. These measurements by the FoCal would constitute an essential part of a comprehensive small-$x$ program at the LHC, with a broad array of complementary probes. While the FoCal has some complementarity to future measurements by the EIC~\cite{AbdulKhalek:2021gbh} and the LHCb experiment, it can reach considerably lower $x$ values than the EIC.

Vector meson photoproduction and electroweak boson studies have provided unique constraints on the gluon parton distribution function for nuclei using Runs 1\& 2 data~\cite{Contreras:2015dqa,Klein:2020fmr, Klein:2020nvu}. They provide clear evidence that the partonic structure of nuclei is different compared to free protons, with nuclear shadowing effects increasing with decreasing longitudinal momentum fractions $x$. In addition, no evidence of saturation of the gluon PDF in the proton is observed between HERA and LHC energies, down to $x\sim 10^{-5}$ \cite{ALICE:2014eof,ALICE:2018oyo}. For a Pb target, the $J/\psi$ photoproduction cross-section is compatible with moderate gluon shadowing, in line with the leading twist approximation and the central values of the EPS09 and EPPS16 nuclear parameterizations~\cite{ALICE:2012yye,ALICE:2013wjo,ALICE:2021gpt,CMS:2016itn}. The ALICE data has smaller uncertainties than these parameterizations, pointing to the utility of UPC data for constraining parton distributions~\cite{Eskola:2022vpi}. The comparison of the energy dependence of photoproduced vector mesons measured in \pPb and \PbPb collisions is expected to significantly reduce the theoretical uncertainties~\cite{Citron:2018lsq}, thus a systematic program of UPC vector meson measurements is needed.

The FoCal will enable measurements of photoproduced J/$\rm \psi$ and $\psi$(2S) mesons to reach a $x\sim 10^{-6}$ and $Q^2\approx3-4$ GeV$^2$. This is a region that cannot be explored by the EIC. The energy dependence of the ratio between $\psi(2S)$ and J/$\rm \psi$ is particularly sensitive to the difference between linear and non-linear gluon evolution~\cite{Hentschinski:2022xnd}, hence saturation effects. Photoproduction data in Runs 3\&4 will allow ALICE to go beyond measurements of vector meson cross-sections. Studies of open charm~\cite{Klein:2002wm} and/or dijets~\cite{ATLAS:2017kwa} are sensitive to gluon distributions with fewer theoretical assumptions than for vector mesons. The study of angular correlations of diffractive dijets~\cite{CMS:2022lbi} is also sensitive to the gluon nuclear density, particularly for low-momentum track-based jets where ALICE would have a competitive advantage.  

Measurements of $d\sigma/dt$ for coherent and incoherent photoproduction of vector mesons are respectively sensitive to the transverse distribution of gluons in a nucleus (similar to a GPD)~\cite{STAR:2017enh, Klein:2021mgd}, and the event-by-event fluctuations in the nuclear configuration, including gluonic hotspots~\cite{Cepila:2016uku,Cepila:2018zky,Klein:2019qfb}. ALICE will collect large enough data samples to make these measurements for a number of different mesons~\cite{Citron:2018lsq}, including the $J/\psi$ and the $\rho$. The new ALICE streaming DAQ system will be important for increasing the vector meson sample sizes, and reducing systematic uncertainties due to the trigger requirements. The FoCal detector will allow ALICE to extend these measurements to even lower $x$ values, providing enhanced sensitivity for gluon saturation. The future UPC measurements using the FoCal will also benefit from having other ALICE detectors. It is anticipated that the Zero Degree Calorimeters can be utilized to distinguish the photon direction in ultra-peripheral \PbPb collisions~\cite{Broz:2019kpl}. This will allow for the exploration of the lowest possible $x$ values in Pb. ALICE measurements can also be carried out combining the FoCal together with the central barrel or the forward muon detector. In this way, a larger kinematic range can be explored, which is vital for gluon saturation searches. More details of the physics prospects in ultra-peripheral heavy-ion collisions using the FoCal detector can be found elsewhere~\cite{Bylinkin:2022temp}.
 
Gluon saturation at low-$x$ has also been postulated to be the source of quantum entanglement effects. These could lead to instant thermalization based on the potential equivalence of entanglement and thermodynamic entropy, and a determination of final state particle production in the context of minimal decoherence effects. In order to study this type of parton-hadron duality, ALICE can complement the EIC measurements in ep and eA systems via measurements in pp, pA and AA. The survival of the coherent state as a function of system size can be measured through particle multiplicities and quantum tomographic correlation functions of particles from low-$x$ processes. Quantum entanglement can also be studied by reconstructing the density state operator using quantum tomography~\cite{Martens:2017cvj}. In this direction, the study of exclusive four $\pi$ photoproduction will provide a search for exited $\rho$ states~\cite{Klusek-Gawenda:2020gwa}.

\item {\it What are the large scale nuclear structures and QGP transport parameters?}

Large scale nuclear structures, such as the nuclear deformation and radius, can be precisely investigated using measurements of anisotropic flow. This is achieved courtesy of the QGP hydrodynamic response, where initial state features are imprinted on final state anisotropic flow observables. These nuclear structures are often accessible via low energy nuclear experiments, but are limited to the electrically charged proton profiles. Measurements in heavy-ion collisions can in principle access the entire nuclear matter profile. These profiles are also inaccessible for DIS nuclear collisions by the virtue of their design to probe sub nucleon scales. Such measurements have already been used to determine $^{129}$Xe deformation, with values of the quadrapole deformation $\beta_{2}=0.18\pm0.02$ obtained in Run 2 from ALICE data~\cite{ALICE:2018lao}. These have never been measured previously. Similar studies have proved equally successful constraining the $^{238}$U and $^{197}$Au deformations from RHIC data~\cite{Giacalone:2021udy}. Firstly, we propose to carry out a suite of anisotropic flow measurements in $^{16}$O--$^{16}$O and p-$^{16}$O collisions. These will occur in Run 3. Hydrodynamic calculations have shown such measurements are sensitive to $\alpha$ clustering in the $^{16}$O nucleus~\cite{Summerfield:2021oex}, a longstanding pursuit in the low energy nuclear community. The clustering effects are calculable in Lattice QCD. Clustering affects anisotropic flow measurements more strongly at LHC energies than at RHIC.

Anisotropic flow and flow fluctuations measurements in central large nuclei A-A collisions are particularly sensitive to the nuclear profile. In addition to Xe--Xe collisions at the LHC, this has also been demonstrated with data from the RHIC Isobar run coupled with hydrodynamic calculations, where these profile parameters have been highly constrained~\cite{Jia:2022qgl}. Anisotropic flow measurements of $v_{2}$ and $v_{3}$ in very central Pb--Pb collisions at the LHC have also revealed the ``ultra-central problem" - that is a deviation from the hydrodynamic description where it is expected to be most applicable (for a recent review see here~\cite{Giannini:2022lbj}). Proposals to explore this further, such as the introduction of an octupole deformation ($\beta_{3}$) for the Pb nucleus ~\cite{Carzon:2020xwp}, which has been predicted for the doubly magic Pb$^{208}$ ground state, require measurements of $v_{3}\{4\}/v_{3}\{2\}$ well beyond the accuracy of those achieved in Runs 1\&2 by ALICE~\cite{ALICE:2011ab}. There are also predictions for an quadrapole deformation of $\beta_{2}=0.05$ for Pb$^{208}$ ~\cite{Raman:2001nnq}, which would lead to finite value of $v_{2}\{4\}$. This has yet to be observed in within statistic precision for previous ALICE data in very central Pb-Pb collisions~\cite{ALICE:2014dwt}. Therefore, the greater statistics provided by Runs 3-6 will be crucial in unraveling these nuclear properties. Their constraints are also critical for the determination of QGP transport parameters obtained by the hydrodynamic framework, in the regime where hydrodynamics is most applicable. These transport parameters include the famous shear viscosity over entropy ratio $\eta/s$. Such nuclear profile constraints could be particularly important for ALICE 3, as non double magic nuclei such as $^{84}$Kr or $^{128}$Xe (with larger deformations compared to the double magic $^{208}$Pb nucleus) are being explored for heavy-ion running.

The hydrodynamic framework used to describe collective flow involves numerous transport parameters, which characterize the coupling of the QGP. To date, only two of those, the shear and bulk viscosity over entropy ratios ($\eta/s$ and $\zeta/s$), have benefited from extensive constraints using heavy-ion data. Those constraints provide evidence the QGP is the most strongly coupled system ever studied in the lab. The corresponding relaxation times, of the order of 1 fm/c or less, demonstrate how thermalization can be achieved on the most rapid timescales ever observed for a many body system. Comparisons of these extracted transport parameters to fundamental descriptions offer a unique test to Holographic models~\cite{Finazzo:2014cna}. These include AdS/CFT, which predicts $\eta/s = 1/4\pi$ in the infinite coupling limit. Such approaches assume a correspondence of 5 dimensional strong gravitational fields in a black hole with a 4 dimensional high temperature QCD system. These transport parameters are also predictable in Lattice QCD, but often have much larger uncertainties~\cite{Ratti:2018ksb}.

Two-particle correlations with net baryons can be used to explore additional transport parameters of the hydrodynamic evolution. The baryon diffusion constant $D_{B}$ is an example of such a parameter beyond $\eta/s$ and $\zeta/s$. It characterizes the mobility of baryon number, and is predicted to be finite at the LHC, despite the fact that $\mu_{B}\sim0$. A two-particle correlation function has been proposed to constrain $D_{B}$~\cite{Floerchinger:2015efa}. It explores correlations of net-baryon fluctuations as a function of azimuthal and rapidity separations. Such an analysis has yet to be carried out from Run 1 or 2 data, since it is statistically challenging. It will be greatly aided by the increase of two orders of magnitude in the \PbPb integrated luminosity foreseen for Runs 3\&4. The ALICE detector is particularly suited to this task, given its world leading particle identification capabilities. Finally, the recently published measurements of balance functions (BFs) of identified charged hadrons $(\pi,\rm K,\rm p)$ in \PbPb collisions from Runs 1\&2 play an important role in constraining the charge diffusion coefficient $D_{e}$ for quarks~\cite{ALICE:2021hjb}. However, the central tracking acceptance of $-1< \eta < 1$ for Runs 1\&2 leads to large uncertainties in $D_{e}$ from ALICE BF data~\cite{Pratt:2021xvg}. In turn, this leads to values between $0.5D_{Latt} <D_{e} < 4D_{Latt}$ being permissible, where $D_{Latt}$ represents the Lattice QCD prediction. These uncertainties will be significantly reduced when the same measurements are performed using the ALICE 3 setup, as the acceptance where identified charged hadron measurements can be made will increase to $-4< \eta < 4$. The reduction in $D_{e}$ uncertainties due to increases in the $\Delta \eta$ coverage are expected to be at least a factor 4.

\item {\it How does the QGP affect hard probes?}

The energetic partons produced in heavy-ion collisions from hard scatterings in the initial collision undergo successive branching. This results in a parton shower that can be modified in the presence of the QGP. While the produced particles are highly collimated about the direction of the initial parton, they also cover a range of different momentum scales. The properties of these collimated sprays of particles, known as jets, and how they emerge from QCD calculations, have been extensively studied in high-energy physics. Jets are a primary tool for uncovering the details of interactions of partons with the QGP medium. This is because jets lose energy and are modified as they traverse the medium, a phenomenon called {\it jet quenching.} ALICE is pursuing a multi-messenger approach that systematically studies different features of jet quenching to explore the microscopic structure of the QGP and its governing degrees of freedom.

The inclusive jet and hadron production, as well as their correlations from low- to high-\pt, allow for an assessment of one of the hallmarks of jet quenching - jet energy loss to the medium. In practice, although energy is conserved, the yields of jets with finite jet resolution parameter $R$ are suppressed in heavy-ion collisions due to jet-medium interactions. The suppression of the leading hadrons and fully reconstructed jets in \PbPb collision at the LHC broadened the observations found at RHIC. That is the QGP is opaque to jets over a large energy range~\cite{ALICE:2010yje, ALICE:2018vuu, ALICE:2019qyj}. Recent measurements of hadron-jet correlations~\cite{RCTaliceQM22} show a potential explanation for the fate of the lost energy in the jet quenching process - in both the momentum and angular scale. The energy is distributed over large angles, and recovered in low-\pt (below 20 GeV/c) jets. Additionally, new results on inclusive jet suppression~\cite{HBaliceQM22}, using Machine Learning techniques~\cite{Haake:2018hqn}, show that larger $R$ jets (up to $R=0.6$) are suppressed relative to smaller $R$ jets at \pt = 40 GeV/c, and the energy is not yet recovered. In Runs 3\&4, ALICE will continue measurements of jet energy loss with inclusive jet suppression and semi-inclusive coincidences of hadron-jet ~\cite{ALICE:2015mdb}, photon-hadron, and photon-jet. The substantial increase in statistics and new experimental techniques, such as mixed events and Machine Learning~\cite{Haake:2018hqn}, will provide increased precision and an extended kinematic range for jets, and high-\pt hadrons and photons.

At the same time, the hadron-jet coincidences are sensitive to the \emph{jet-correlated} response of the medium to the presence of energetic probes~\cite{ALICE:2015mdb}. This flux of medium energy induced by the propagating jet is one of the most sought after consequences of jet-medium interactions~\cite{Cao:2020wlm}. A medium composed of scattering centers - quasi-particles - ought to, with some finite probability, induce large angle Moliere scatterings of the propagating partons ~\cite{DEramo:2018eoy,Barata:2020rdn}. Such a jet deflection should manifest itself in the modified acoplanarity of the di-jet pair in \PbPb compared to \pp collisions. Utilizing hadron-jet coincidences, ALICE has investigated this effect at the lowest jet \pt possible and with a large jet $R$~\cite{RCTaliceQM22}. Additionally, jet deflection should manifest itself in large momentum kicks to the hard core of the jet~\cite{Barata:2021wuf, DEramo:2018eoy}. ALICE also uses the groomed hardest $k_{T}$ to probe this by looking at differences between \pp and \PbPb collisions at larger groomed $k_{\rm T}$ values~\cite{HBaliceQM22}. Likewise, in this area, Runs 3\&4 will provide further precision to put improved quantitative constraints on the probability of this process within the QGP.

Jet fragmentation patterns and modification of the jet structure are also key to understanding of the interactions of the medium with the jet. ALICE has for the first time measured fully corrected angular and momentum sub-jet structure of jets~\cite{ALargeIonColliderExperiment:2021mqf,ALICE:2022vsz}, and has made advances towards measurements of other quantities such as jet angularities, jet-axes, the Lund Plane, and intrajet hadron correlations ~\cite{JMaliceQM22,RCTaliceQM22,ALICE:2022hyz, ALICE:2021yet}. These measurements are compared to both Monte Carlo models and analytical calculations. They offer a new and more stringent view of the jet-medium interactions. In particular, new measurements show the largest angle splittings within a groomed jet are suppressed as compared to small-angle splittings, which can be connected to a characteristic aperture scale at which the jet fragments/prongs interact with the medium incoherently. Moreover, the new measurements of leading subjets suggest sensitivity of the predicted flavor dependence of jet quenching that should be quantified with more precise data. Concurrently, the new measurements of jet-axes disfavor the intra-jet \pt broadening, as prescribed by the BDMPS formalism as the main mechanism of energy loss in the QGP~\cite{Ringer:2019rfk}. The Run 3\&4 data will be instrumental in providing the necessary precision to further investigate such observations. However, most importantly, the new high statistics data will enable new differential measurements~\cite{Citron:2018lsq}. These will  study different regions of the parton shower phase-space related to rare/hard vs. multiple-soft medium-induced radiation. This can be achieved via simultaneous measurements of angular and momentum distributions of jet structure such as the Jet Lund Plane~\cite{Lifson:2020gua}.

The modification of jets in the medium is expected to depend on the path the jet traveled in the medium. ALICE traditionally measures this in two ways. They involve using correlations and measuring the jet yield suppression with respect to the event-plane angle. Measurements by ALICE of hadron-jet correlations show no dependence on the angle of the jet with respect to the event plane angle within the uncertainties~\cite{ALICE:2019sqi}. On the other hand, measurements of the jet $v_{2}$ demonstrate a significant azimuthal anisotropy~\cite{ALICE:2015efi}. ALICE has continued to explore these effects using higher statistics data from Run 2. For example, a technique that utilizes Event Shape Engineering~\cite{Schukraft:2012ah} to select events based on their anisotropy within a centrality class, demonstrates an increased suppression of out-of-plane yields for more anisotropic events~\cite{CBaliceQM22}. This technique can be further pursued using the higher statistics Run 3\&4 data to investigate the path length dependence of jet substructure modification. Finally, a first look at photon-jet correlation measurements with ALICE was investigated~\cite{ALaliceQM22}. These studies demonstrated the ability to measure photon-tagged jets at lower $p_{\rm T}^{\gamma}$ (20~\GeVc). This measurement is statistically limited, thus future ALICE photon-jet measurements will be crucial for measuring the absolute energy loss, since the photon provides an estimate of the initial \pt of the jet. 

ALICE has also performed measurements of the heavy-flavor hadrons as an additional probe of parton-medium interactions, with examples found in ~\cite{ALICE:2021rxa, ALICE:2022tji, ALICE:2019nuy}. Heavy flavor hadrons offer a unique opportunity to study the medium effects because of the well established probe - heavy quarks. These are created early in the hard scatterings, and maintain their flavor identity throughout the evolution of the medium. A new set of measurements focusing on the leading heavy-flavor particles will take advantage of this theoretical control. It will also profit from improved experimental control, such as better discrimination of "combinatorial jets" - the uncorrelated background to hard scattering flux of energy, over the light flavor measurements. This applies to both measurements of the heavy-flavor jet \pt spectra and jet substructure observables. One of the striking findings by ALICE is the observed mass dependence of energy loss. This manifests itself in smaller nuclear modification factors for prompt $D$ mesons (formed from charm) compared to non-prompt $D$ mesons (formed from bottom quarks) at \pt $> 8$ \GeVc. This onset is interpreted as the dead cone effect~\cite{ALICE:2021aqk}. In future runs, ALICE will take advantage of its precision tracking to study this effect in more detail, and contrast it with the impact of gluon splitting processes within the parton shower~\cite{Voutilainen:2015lqa}.

In general, with the newly upgraded ALICE detector, the data from Runs3\&4 are expected to improve the quantification of how light and heavy quarks propagate within the medium. The new high-statistics measurements will bring additional constraints to the extraction of the transverse and longitudinal transport coefficients. These include the jet transport parameter $\hat{q}$~\cite{JETSCAPE:2021ehl,JETSCAPE:2022ixz}, which quantifies the transverse diffusion. The longitudinal drag $\hat{e}$ and diffusion $\hat{e_{2}}$ coefficients~\cite{Cao:2017crw}, as well as the heavy-flavor diffusion coefficient $D_{s}$~\cite{Moore:2004tg}, are also expected to be better constrained. These coefficients characterize the coupling of such processes to the medium, and the ultimate task is to perform a dedicated set of measurements that will allow further connections to the emergence of the QGP on a microscopic level. Runs 3\&4 will also allow for a revisiting of the jet substructure measurements for the heavy-flavor induced jets e.g. angular structure of groomed and ungroomed jets. This can be achieved over a broad kinematic regime, that is largely unique to ALICE (from \pt$\approx~0$ of heavy-flavor hadrons to hundreds of \GeVc\ jets). The new data will enable ALICE to fully explore the momentum and the angular correlations within the jet substructure with a systematic study of the Lund Plane (see considerations in~\cite{Citron:2018lsq}). With this new data, and new techniques, the flavor dependence of the Lund Plane can be extended beyond the charm sector~\cite{ALICE:2021aqk} to the beauty sector~\cite{Dreyer:2021hhr}. 

For Runs 5\&6, ALICE 3 will provide an environment for precision jet measurements over much larger kinematic ranges than previously accessible at the LHC. Specifically, jet hadrochemistry will become possible over such a range with the excellent PID capabilities for identifying hadrons inside jets. This is valuable to study in vacuum to provide inputs to hadronization models~\cite{ATLAS:2022zgo}, and also in the QGP since the interactions with the medium could change the chemistry make-up of a jet~\cite{Sapeta:2007ad}. Additionally, ALICE 3 will provide increased resolution, purity, and statistics for heavy-flavor jets. The increased psuedorapidity coverage will allow for measurements of heavy-flavor jet correlations or photon-heavy-flavor jet correlations with unprecedented purity at low transverse momentum scales, providing insight into the microscopic properties of in-medium energy loss, including its path length dependence. Finally, ALICE 3 allows for the possibility to resolve heavy-flavor hadron pairs inside of jets to probe gluon splittings, which can be used to study the space-time picture of jet quenching~\cite{Attems:2022ubu}, and access a pure gluon sample of jets.

Finally, while we are in the midst of fully understanding the measurements that need to be done, the continued development of analysis techniques is essential for jets in a complex environment. This is related to both the careful selection of the most useful measurements~\cite{Lai:2021ckt}, but also to the development of the appropriate correction strategies allowing for meaningful theoretical comparisons. These include fully corrected substructure measurements~\cite{Andreassen:2019cjw,Mulligan:2020tim}, and background mitigated hadron-jet correlations down to lowest jet momenta with large jet resolution parameters. At the same time, it is important to encourage the divulgence of uncertainty correlations for the new generation of measurements. Such an activity will take advantage of the newly developed Bayesian inference methods~\cite{JETSCAPE:2021ehl,Nijs:2020roc}. This will also aid Machine Learning projects that will drive the development of the modeling of jet-medium interactions in the coming years~\cite{Chien:2018dfn,Apolinario:2021olp,Lai:2021ckt,Liu:2022hzd}.

\item {\it How do rarely produced hadrons form and interact?}

A detailed program by ALICE for Runs 3-6 will shed further light on more exotic heavier flavor states in order to unambiguously determine the hadronization mechanisms, and thus the creation of matter, as a function of flavor. In this regard, the measurements of multi-heavy-flavour hadrons and exotic states offer unprecedented sensitivity. To achieve high statistics for multi-heavy-flavour particles, a novel experimental approach is needed to track all their decay products, typically including hyperons, before they decay. This calls very high tracking/vertexing precision very close to the interaction point, particle identification capabilities over a wide transverse momentum range, and high readout rates. Moreover, large acceptance is required, not only for reasons of statistics, but also in order to investigate the dependence of the production of multi-heavy-flavour hadrons on the variation of the heavy quark density with rapidity.

Recent flavor dependent hadronization measurements have become a unifying theme for the bulk and jet sector of QCD. The hadronization of a final state parton cannot be described adopting a perturbative approach, and is usually modelled through a phase of string breaking and/or cluster formation, which is considered to be independent of the surrounding parton density~\cite{Andersson:1983ia}. However, as also confirmed by measurements of e.g. baryon yields in pp collisions at the LHC ~\cite{ALICE:2016fzo,ALICE:2020wfu}, additional hadronization mechanisms may exist, whereby quarks that are close in phase space can combine into colourless hadrons. Dynamic modeling of these processes, in particular in the strange and charm sector, has led to many novel approaches that were implemented in established event generator models, such as PYTHIA. Color reconnection (CR) and multiple parton interactions (MPI) were attempts to parameterize features of the initial state and final state into the actual formation process of hadrons. Furthermore, it was shown that initial state gluon saturation might lead to quantum entanglement effects, which impact the final state hadron multiplicities~\cite{Tu:2019ouv}. 

Experimental advances in the strange sector include the detailed mapping of strangeness enhancement in proton-proton collisions as a function of centrality~\cite{ALICE:2016fzo}, and the potential flavor hierarchy in the light/strange hadron formation during the QCD crossover~\cite{Flor:2020fdw}. These led to refinements of the transport codes that were then followed up by more detailed measurements of baryon to meson formation in the strange sector. Such advancements have been studied in the charm sector, as a function of system size~\cite{ALICE:2020wfu}. In heavy-ion collisions, where partons may travel freely over distances much larger than the typical hadron sizes, and a dense system of partons close to thermal equilibrium is formed, recombination mechanisms become more dominant. These make the production of baryons and other heavy hadrons more favourable than in pp collisions~\cite{ALICE:2021bib}. First measurements of baryon/meson ratios in the charm sector also indicate a low-\pt enhancement consistent with such a picture that has dominated similar measurements in the light and strange sector associated with the soft underlying event~\cite{ALICE:2021vxl}. 

A comprehensive campaign of precision measurements of charm baryon production yields across collision systems is planned for Runs 3\&4. This starts with baryons containing only one charm quark, but several light and strange quark combinations. The evolution of particle production as a function of increasing strangeness or charm content is one of the main open questions regarding hadron formation, which is a key feature of the non-perturbative sector of QCD. For example, LHCb has found a series of tetra- and penta-quark states in the charm sector. A detailed study of strange penta- and hexa-quark states in ALICE found no evidence in the strange sector. The measured upper limits are several orders of magnitude below estimates from statistical hadronization models (SHM), which have done extremely well on postulating the yields of composite objects in the light sector~\cite{Andronic:2017pug}. SHM predictions assume full equilibration of all flavors up to charm. The comparison of the measured yields to these predictions would provide a very sensitive measure of the degree of equilibration of charm quarks in the medium, including its system-size dependence. 

Ultimately, it must be possible to relate the equilibration level to the transport properties of the QGP via kinetic transport modeling in order to provide a consistent description of the entire phenomenology. Complementing these opportunities, relativistic heavy-ion collisions provide the only available tool to study the formation and interaction of exotic hadronic states inside a medium filled with deconfined color charges. This puts new constraints on the properties of the states, the characterization of their binding potential, and the details of their hadronization mechanisms. Nuclear collisions at the LHC energies, in particular, offer a unique opportunity to test these mechanisms in the presence of large charm-quark densities. In this environment, while the screening of color charges may reduce the yield of certain bound states, medium-induced or medium enhanced mechanisms may enhance their production in the low-\pt region, as predicted by statistical hadronization models. The interest of these observations is twofold. On one hand, the rates of formation and dissociation of bound states depend on their binding energy and size, and the measurement of the production rates and anisotropic flow will allow one to distinguish compact multi-quark configurations from molecular states. On the other hand, states whose nature can be determined by other means, for example by measuring interaction potentials through final state interactions, could be exploited as a calibrated probe in nuclear collisions. This would shed light on the underlying mechanisms and timescales driving the hadronization in a deconfined medium, and test the properties of the deconfined medium itself.

Predictions for light flavor tetraquark states have existed for over 40 years~\cite{Jaffe:1976ig}, but the existence of these states has up to now not been confirmed. To begin to address this, ALICE has published three papers on tetraquark studies of the $a_0(980)$ resonance, a tetraquark  candidate, using kaon-kaon correlations from Runs 1\&2~\cite{ALICE:2017jto,ALICE:2018nnl,ALICE:2021ovd} for \pp and \PbPb collisions. In Runs 3\&4, ALICE will continue these two-meson correlation studies for other low lying tetraquark candidates, the $K^*_0(700)$ and $f_0(500)$. The high rate capabilities in Run 3\&4, will also allow access to rare phenomena in the hadronic phase formed after the phase transition. These include the production of nuclei, hypernuclei and, combined with excellent heavy-flavour capabilities, possibly the as-yet undiscovered supernuclei (nuclei in which a nucleon is replaced by a charmed baryon). Wide acceptance, again combined with outstanding heavy-flavour performance at low transverse momenta, will allow to extend the studies of hadron-hadron potentials via two-particle correlations to the charm sector. This will provide a powerful tool to investigate the structure of the newly discovered charmed exotic states. In Runs 5\&6, the measurement of the production yields of multi-charm baryons, which can only be produced by combination of uncorrelated charm quarks, would provide a qualitatively new handle on the production of heavy-flavour hadrons. Measurements of multi-charm hadrons, such as the $\Xi^{+}_{cc}$ (ccd), $\Xi^{++}_{cc}$ (ccu), $\Omega^{+}_{cc}$ (ccs), $\Omega^{++}_{ccc}$ (ccc), and exotic states such as the newly discovered T$^{+}_{cc}$ (ccud)~\cite{LHCb:2021vvq,LHCb:2021auc}, would provide a direct window on hadron formation from the QGP. In fact, the yields of multi-charm baryons relative to the number of produced charm quarks are predicted to be significantly enhanced in AA relative to pp collisions ~\cite{Andronic:2021erx}. Enhancements are expected by as much as a factor 100 for the recently observed $\Xi_{cc}$ baryon~\cite{LHCb:2017iph}, and even by as much as a factor 1000 for the as yet undiscovered $\Omega_{ccc}$ baryon. The observation and precise quantification of such effects, would represent a major discovery for the study of the properties of deconfined matter. 
    
In addition, LHC heavy-ion collisions also have a discovery potential for more exotic hadronic states. This is because of the long-lived deconfined medium and the large cross sections for heavy quarks available at the highest energies. These provide an ideal playground for the production of exotic states with e.g. multiple b quarks. The most notable case is the T$^{-}_{bb}$ bb, whose experimental detection would profit from the long predicted lifetime (c$\tau$ $\sim$ 2.3 mm) due to the stability of the state with respect to strong decays~\cite{Hernandez:2019eox} (contrary to the recently discovered, shortly-lived T$^{+}_{cc}$ state) . Discovery opportunities also exist for compact bound hidden-charm hexaquarks, also predicted to be stable with respect to strong interaction decays~\cite{Liu:2021gva}, and molecular states composed of three D mesons. Studying exotic QCD states in nuclear collisions is therefore of central importance for QCD physics. The feasibility of such studies has been demonstrated by a recent measurement of $\chi_{c1}$(3872) production in Pb–-Pb collisions by the CMS collaboration in the range \pt $>$ 10 GeV/c ~\cite{CMS:2021znk}. An immediate goal for ALICE 3 is to measure the production of $\chi_{c1}$(3872) down to the \pt region $<$5-6 GeV/c, where a significant enhancement of the yield was predicted~\cite{Andronic:2019wva}. This is not accessible by other LHC experiments. For hadrons containing charm and beauty quarks, scattering experiments are not feasible to determine the interaction strength between hadrons, therefore the only way to access the information is the femtoscopy technique. This technique consists of the measurement of correlations in momentum space for hadron-hadron pairs, and can be used to extract the corresponding scattering parameters~\cite{ALICE:2020mfd}. The ALICE 3 upgrade will allow the measurement of several hadron combinations including DD$^{*}$, $\Lambda_{c}^{+}\Sigma_{c}^{0;+;++}$ and BB$^{*}$ in pp, p–Pb and Pb–Pb collisions and thereby shed light on the nature of many exotic hadrons. In particular, using the same method as mentioned above in studying tetraquark states with meson-meson correlations, the ALICE 3 upgrade will allow the study of the $f_2(2010)$ and $\psi(3770)$ tetraquark candidates with $\phi\phi$ and $D^+D^-$ correlations, respectively.

\item {\it What are the mechanisms that lead to QGP-like signals in small systems?}

The formation of the QGP in heavy-ions has been confirmed via numerous measurements. A key example is the simultaneous observation of collective flow phenomena traced to a strongly coupled liquid, and jet quenching. Measurements in small collision systems, such as \pp and \pPb at the LHC, have shown particle correlations and strangeness yields in high-multiplicity collisions resemble observations associated with the creation of the QGP~\cite{ALICE:2012eyl,ALICE:2013snk,ALICE:2015lpx,ALICE:2016fzo}. On the other hand, jet quenching in small systems has not been observed. The resolution of this enigma remains one of the most important questions for our field. ALICE has contributed with a stringent limit on energy loss within small collision systems~\cite{ALICE:2017svf}, and plans to improve on those measurements with the high-statistics data in the upcoming LHC runs~\cite{Citron:2018lsq}. ALICE can also use machine learning methods to design and measure new jet observables that are maximally modified by traversing small systems~\cite{Lai:2021ckt,Lee:2022kdn}.

Anisotropic flow measurements in high-multiplicity small system collisions at RHIC and the LHC are consistent with hydrodynamic predictions~\cite{PHENIX:2018lia,Schenke:2020mbo}. This indicates these collisions may create the smallest possible QGP droplets. However, these measurements from Runs 1\&2 suffer from large uncertainties due to ambiguities in the non-flow subtraction methods pursued at the LHC~\cite{Lim:2019cys}. For Run 4, the introduction of the FoCal will enable $\Delta \eta$ separations of up to 9 units for two-particle correlation anisotropic flow measurements of $v_n$, with such separations being critical in reducing non-flow. The broad increase from the ALICE 3 acceptance will also reduce non-flow contributions in two-particle correlation $v_n$ measurements by a factor of 4 for Runs 5\&6~\cite{Voloshin:2008dg}. Four particle $v_n$ measurements enjoy a reduction of a factor 64, and this increases exponentially with the number of particles used to determine $v_n$. Such reductions are vital in the transition of anisotropic flow measurements in small systems from being qualitative to truly quantitative. Measurements in Run 4 and ALICE 3 are therefore essential for precise quantitative constraints of the studies of the applicability of hydrodynamics at its limits in small systems~\cite{Nagle:2018nvi}. The improvements in non-flow suppression and increased statistics are also essential for identified hadron $v_n$ measurements vs. \pt in small systems. These test hydrodynamic mass ordering at low-\pt, with arises due to a common radial flow profile. Such measurements, using ALICE's highly competitive particle identification capacities, will be key for all future precision tests of hydrodynamics in small systems. These tests can also be expanded for the unmeasured $\Xi$ and $\Omega$ $v_n$ vs. \pt, which have proved statistically challenging in Runs 1\&2.

Finally, the increased statistics for extremely high multiplicity \pp collisions in Runs 3\&4 will allow for measurements of the $\Omega/\pi$ ratio at $\rm{d}N_{ch}/\rm{d}\eta \sim 96$. This is six times larger than corresponding measurements from Runs 1\&2, and fully overlap with ALICE's \PbPb measurements~\cite{ALICE:2016fzo}. This will in turn allow for a critical test regarding particle production mechanisms and the interpretation of strangeness enhancment in \pp collisions~\cite{Citron:2018lsq}. If this ratio saturates and reaches the thermal limit in \PbPb collisions, this would support the idea the QGP can be created in \pp collisions. On the other hand, if this ratio continues to increase beyond the \PbPb values, this perhaps would favor non-QGP mechanisms, such as color ropes implemented in the DIPSY model~\cite{Bierlich:2015rha}, that predict a continuing increase.

\item {\it What are the connections and broad impacts of ALICE measurements to other fields of physics?}

One of the recent major discoveries are gravitational wave signals from neutron star mergers. The tidal deformation during approach and the ring-down of the frequency spectrum after the merger hold potentials clues on the core composition of neutron stars. In particular, this applies for stars with a large mass to radius ratio. Model predictions range from simple neutron matter, to hyperon matter, and deconfined quark matter~\cite{Annala:2019puf}. Theoretically, the main required ingredient is an equation of state for the system as a function of density and temperature. Experimentally, relativistic heavy-ion experiments, and in particular ALICE with its high luminosity and data rate beyond Run 3, will be able to map out the production of hyperons, hypermatter and quark matter. Since deconfined matter and hyperon production have been discussed in the previously, the focus here is on hyper nuclei production and the measurement of the balance of attractive and repulsive forces in the interaction between hyperons and protons/neutrons through femtoscopic measurements. During Runs 1\&2 these studies have shown great promise as documented in a recent Nature article by ALICE~\cite{ALICE:2020mfd}. Certain di-baryon correlations in the light and strange sector show attractive potentials, which should lead to bound hexa-quark states. These results should also be viewed in the context of hyper nuclei production. In previous campaigns, ALICE has successfully reconstructed the lightest hypernucleus, the hypertriton and its anti-particle ~\cite{ALargeIonColliderExperiment:2021puh}.  

In Run 4, we hope to extend the statistical sample in \PbPb collisions to a level that might make the reconstruction of A=4 hypernuclei, in particular the $^{4}_{\Lambda}$H. Their measurement is interesting for precision tests of particle production models~\cite{Bellini:2019zqc} and to constrain hyperon-nucleon potentials~\cite{Haidenbauer:2019thx}, and thus the formation probability of hyper-matter in dense stellar objects. Regarding Runs 5\&6, anti-nuclei and anti-hyper-nuclei with A$>$4 such as anti-$^{5}_{\Lambda}$He or anti-$^{6}$Li have yet to be discovered, and may well be in reach of ALICE 3. The ALICE 3 apparatus is ideally suited for the observation of the $A=4$ or $A=5$ hyper-nuclei like $^{4}_{\Lambda}$H or $^{5}_{\Lambda}$He. The measurement of $A=6$ nuclei would provide precision tests for the formation of bound clusters thanks to the special nature of $^{6}$He and $^{6}$Li. $^{6}$He is the lightest known (anti-)halo-nucleus. Its production is therefore expected to be suppressed in coalescence models with respect to thermal-statistical models, due to its much larger size. $^{6}$ Li is a stable isotope with a spin of $J=1$. With respect to the helium isotopes $^{4}$He and $^{6}$He with $J=0$, $^{6}$Li production is therefore expected to be enhanced by the degeneracy factor g of its spin-substates; $g = 2J +1 = 3$. The expected production yields $d$N/$dy$ for A=5 hyper-nuclei and A=6 nuclei are in the 10$^{-9}$ to 10$^{-11}$ range in \PbPb collisions. 

\item {\it What can ALICE achieve regarding Beyond Standard Model (BSM) Physics?}

Although studying BSM physics is not ALICE's primary purpose, there are areas where it can make extremely useful contributions, with the benefit of the upgrades and increased statistics in future runs. One such area is the study of light-by-light scattering, $\gamma\gamma\rightarrow\gamma\gamma$, using UPCs. Light-by-light scattering occurs via a box diagram, which includes contributions from all (standard model or BSM) electrically charged particles.  In addition, it can be mediated by axion-like particles, $\gamma\gamma\rightarrow A\rightarrow\gamma\gamma$, which leads to a diphoton invariant mass spectrum that is peaked at the axion mass.  By virtue of its sensitivity to lower \pt particles, ALICE (especially ALICE 3) can probe lower axion masses than ATLAS or CMS.  ALICE is also sensitive to two-photon production of $\tau^+\tau^-$. Such studies of this reaction can be used to put limits on BSM phenomena such as a $\tau$ anomalous magnetic moment. 

The discovery of stable massive particles (SMPs) beyond the Standard Model (BSM), such as monopoles, gluinos, heavy leptons, etc., would address a number of important questions in modern physics, including the nature of dark matter, and the unification of fundamental forces. The majority of predictions suggest that SMPs are far too massive to be produced in a foreseeable accelerator~\cite{Fairbairn:2006gg}. However, there are suggestions some of them, e.g. magnetic monopoles, could appear in a mass range accessible at the LHC~\cite{Bruce:2018yzs}, in particular in \PbPb collisions. Detection of such particles is a very challenging due to their highly ionizing nature, which leads to saturation and even malfunction of electronics of e.g. silicon-based detectors~\cite{CMSTracker:2005iuk}. The newly upgraded GEM (Gas Electron Multiplier) based Time Projection Chamber (TPC) of ALICE is free from this challenge, and is capable of measuring very large energy deposits anticipated from monopoles. 

An exciting avenue to study dark matter signals lies with cosmic-ray antinuclei, such as antihelium. This is considered another promising signature of the existence of weakly-interactive mass particles (WIMP), which represent an important candidate for dark matter~\cite{Donato:1999gy}. Since the background from hadronic interactions of primary rays are negligible~\cite{Duperray:2005si}, the preliminary evidence of a handful of anti-$^{3}$He events collected by the AMS collaboration, could be due to a previously neglected process. Namely, these could be the production of $\Lambda_{b}^{0}$ baryons in dark-matter annihilation, and their subsequent decay into anti-$^{3}$He \cite{Winkler:2020ltd}. However, the decay rates of $\Lambda_{b}^{0}$ to anti-nuclei are not experimentally measured, and serve as crucial inputs to understand the AMS data. The large beauty-quark production cross section at LHC energies, and the identification capabilities for nuclei over practically the full phase space, make ALICE 3 the ideal experiment for such studies.

\end{enumerate}
\newpage
\bibliographystyle{utphys}   
\bibliography{ref}

\providecommand{\href}[2]{#2}\begingroup\raggedright\begin{thebibliography}{100}

\bibitem{ALICE-PUBLIC-2019-001}
{\bfseries ALICE} Collaboration, ``{ALICE upgrade physics performance studies
  for 2018 Report on HL/HE-LHC physics}'',.
  \url{https://cds.cern.ch/record/2661798}.

\bibitem{ALICE-PUBLIC-2019-005}
{\bfseries ALICE} Collaboration, ``{A Forward Calorimeter (FoCal) in the ALICE
  experiment}'', Tech. Rep. ALICE-PUBLIC-2019-005, Oct, 2019.
\newblock \url{https://cds.cern.ch/record/2696471}.

\bibitem{ALICE:2022wwr}
{\bfseries ALICE} Collaboration, ``{Letter of intent for ALICE 3: A
  next-generation heavy-ion experiment at the LHC}'',
  \href{http://arxiv.org/abs/2211.02491}{{\ttfamily arXiv:2211.02491
  [physics.ins-det]}}.

\bibitem{ALICE:2803563}
{\bfseries ALICE} Collaboration, ``{Letter of intent for ALICE 3: A next
  generation heavy-ion experiment at the LHC}'', tech. rep., CERN, Geneva,
  2022.
\newblock \url{https://cds.cern.ch/record/2803563}.

\bibitem{ALICE:2022qhn}
{\bfseries ALICE} Collaboration, ``{Performance of the ALICE Electromagnetic
  Calorimeter}'', \href{http://arxiv.org/abs/2209.04216}{{\ttfamily
  arXiv:2209.04216 [physics.ins-det]}}.

\bibitem{ALICE:2021aqk}
{\bfseries ALICE} Collaboration, S.~Acharya {\em et~al.}, ``{Direct observation
  of the dead-cone effect in quantum chromodynamics}'',
  \href{http://dx.doi.org/10.1038/s41586-022-04572-w}{{\em Nature} {\bfseries
  605} no.~7910, (2022) }, \href{http://arxiv.org/abs/2106.05713}{{\ttfamily
  arXiv:2106.05713 [nucl-ex]}}. [Erratum: Nature 607, E22 (2022)].

\bibitem{ALICE:2016fzo}
{\bfseries ALICE} Collaboration, J.~Adam {\em et~al.}, ``{Enhanced production
  of multi-strange hadrons in high-multiplicity proton-proton collisions}'',
  \href{http://dx.doi.org/10.1038/nphys4111}{{\em Nature Phys.} {\bfseries 13}
  (2017) }, \href{http://arxiv.org/abs/1606.07424}{{\ttfamily arXiv:1606.07424
  [nucl-ex]}}.

\bibitem{ALICE:2010suc}
{\bfseries ALICE} Collaboration, K.~Aamodt {\em et~al.}, ``{Elliptic flow of
  charged particles in Pb-Pb collisions at 2.76 TeV}'',
  \href{http://dx.doi.org/10.1103/PhysRevLett.105.252302}{{\em Phys. Rev.
  Lett.} {\bfseries 105} (2010) },
  \href{http://arxiv.org/abs/1011.3914}{{\ttfamily arXiv:1011.3914 [nucl-ex]}}.

\bibitem{ALICE:2013snk}
{\bfseries ALICE} Collaboration, B.~B. Abelev {\em et~al.}, ``{Long-range
  angular correlations of $\rm \pi$, K and p in p-Pb collisions at
  $\sqrt{s_{\rm NN}}$ = 5.02 TeV}'',
  \href{http://dx.doi.org/10.1016/j.physletb.2013.08.024}{{\em Phys. Lett. B}
  {\bfseries 726} (2013) }, \href{http://arxiv.org/abs/1307.3237}{{\ttfamily
  arXiv:1307.3237 [nucl-ex]}}.

\bibitem{ALICE:2010mlf}
{\bfseries ALICE} Collaboration, K.~Aamodt {\em et~al.}, ``{Centrality
  dependence of the charged-particle multiplicity density at mid-rapidity in
  Pb-Pb collisions at $\sqrt{s_{NN}}=2.76$ TeV}'',
  \href{http://dx.doi.org/10.1103/PhysRevLett.106.032301}{{\em Phys. Rev.
  Lett.} {\bfseries 106} (2011) },
  \href{http://arxiv.org/abs/1012.1657}{{\ttfamily arXiv:1012.1657 [nucl-ex]}}.

\bibitem{ALICE:2011ab}
{\bfseries ALICE} Collaboration, K.~Aamodt {\em et~al.}, ``{Higher harmonic
  anisotropic flow measurements of charged particles in Pb-Pb collisions at
  $\sqrt{s_{NN}}$=2.76 TeV}'',
  \href{http://dx.doi.org/10.1103/PhysRevLett.107.032301}{{\em Phys. Rev.
  Lett.} {\bfseries 107} (2011) },
  \href{http://arxiv.org/abs/1105.3865}{{\ttfamily arXiv:1105.3865 [nucl-ex]}}.

\bibitem{ALICE:2022wpn}
{\bfseries ALICE} Collaboration, ``{The ALICE experiment - A journey through
  QCD}'', \href{http://arxiv.org/abs/2211.04384}{{\ttfamily arXiv:2211.04384
  [nucl-ex]}}.

\bibitem{AbdulKhalek:2021gbh}
R.~Abdul~Khalek {\em et~al.}, ``{Science Requirements and Detector Concepts for
  the Electron-Ion Collider}: {EIC Yellow Report}'',
  \href{http://dx.doi.org/10.1016/j.nuclphysa.2022.122447}{{\em Nucl. Phys. A}
  {\bfseries 1026} (2022) }, \href{http://arxiv.org/abs/2103.05419}{{\ttfamily
  arXiv:2103.05419 [physics.ins-det]}}.

\bibitem{Contreras:2015dqa}
J.~G. Contreras and J.~D. Tapia~Takaki, ``{Ultra-peripheral heavy-ion
  collisions at the LHC}'',
  \href{http://dx.doi.org/10.1142/S0217751X15420129}{{\em Int. J. Mod. Phys. A}
  {\bfseries 30} (2015) }.

\bibitem{Klein:2020fmr}
S.~Klein and P.~Steinberg, ``{Photonuclear and Two-photon Interactions at
  High-Energy Nuclear Colliders}'',
  \href{http://dx.doi.org/10.1146/annurev-nucl-030320-033923}{{\em Ann. Rev.
  Nucl. Part. Sci.} {\bfseries 70} (2020) },
  \href{http://arxiv.org/abs/2005.01872}{{\ttfamily arXiv:2005.01872
  [nucl-ex]}}.

\bibitem{Klein:2020nvu}
S.~Klein {\em et~al.}, ``{New opportunities at the photon energy frontier}'',
  \href{http://arxiv.org/abs/2009.03838}{{\ttfamily arXiv:2009.03838
  [hep-ph]}}.

\bibitem{ALICE:2014eof}
{\bfseries ALICE} Collaboration, B.~B. Abelev {\em et~al.}, ``{Exclusive
  $\mathrm{J/}\psi$ photoproduction off protons in ultra-peripheral p-Pb
  collisions at $\sqrt{s_{\rm NN}}=5.02$ TeV}'',
  \href{http://dx.doi.org/10.1103/PhysRevLett.113.232504}{{\em Phys. Rev.
  Lett.} {\bfseries 113} no.~23, (2014) },
  \href{http://arxiv.org/abs/1406.7819}{{\ttfamily arXiv:1406.7819 [nucl-ex]}}.

\bibitem{ALICE:2018oyo}
{\bfseries ALICE} Collaboration, S.~Acharya {\em et~al.}, ``{Energy dependence
  of exclusive $\mathrm {J}/\psi $ photoproduction off protons in
  ultra-peripheral p\textendash{}Pb collisions at $\sqrt{s_{\mathrm
  {\scriptscriptstyle NN}}} = 5.02$ TeV}'',
  \href{http://dx.doi.org/10.1140/epjc/s10052-019-6816-2}{{\em Eur. Phys. J. C}
  {\bfseries 79} no.~5, (2019) },
  \href{http://arxiv.org/abs/1809.03235}{{\ttfamily arXiv:1809.03235
  [nucl-ex]}}.

\bibitem{ALICE:2012yye}
{\bfseries ALICE} Collaboration, B.~Abelev {\em et~al.}, ``{Coherent $J/\psi$
  photoproduction in ultra-peripheral Pb-Pb collisions at $\sqrt{s_{NN}} =
  2.76$ TeV}'', \href{http://dx.doi.org/10.1016/j.physletb.2012.11.059}{{\em
  Phys. Lett. B} {\bfseries 718} (2013) },
  \href{http://arxiv.org/abs/1209.3715}{{\ttfamily arXiv:1209.3715 [nucl-ex]}}.

\bibitem{ALICE:2013wjo}
{\bfseries ALICE} Collaboration, E.~Abbas {\em et~al.}, ``{Charmonium and
  $e^+e^-$ pair photoproduction at mid-rapidity in ultra-peripheral Pb-Pb
  collisions at $\sqrt{s_{\rm NN}}$=2.76 TeV}'',
  \href{http://dx.doi.org/10.1140/epjc/s10052-013-2617-1}{{\em Eur. Phys. J. C}
  {\bfseries 73} no.~11, (2013) },
  \href{http://arxiv.org/abs/1305.1467}{{\ttfamily arXiv:1305.1467 [nucl-ex]}}.

\bibitem{ALICE:2021gpt}
{\bfseries ALICE} Collaboration, S.~Acharya {\em et~al.}, ``{Coherent $J/\psi$
  and $\psi'$ photoproduction at midrapidity in ultra-peripheral Pb-Pb
  collisions at $\sqrt{s_{\mathrm{NN}}}~=~5.02$ TeV}'',
  \href{http://dx.doi.org/10.1140/epjc/s10052-021-09437-6}{{\em Eur. Phys. J.
  C} {\bfseries 81} no.~8, (2021) },
  \href{http://arxiv.org/abs/2101.04577}{{\ttfamily arXiv:2101.04577
  [nucl-ex]}}.

\bibitem{CMS:2016itn}
{\bfseries CMS} Collaboration, V.~Khachatryan {\em et~al.}, ``{Coherent
  $J/\psi$ photoproduction in ultra-peripheral PbPb collisions at $\sqrt
  {s_{NN}} =$ 2.76 TeV with the CMS experiment}'',
  \href{http://dx.doi.org/10.1016/j.physletb.2017.07.001}{{\em Phys. Lett. B}
  {\bfseries 772} (2017) }, \href{http://arxiv.org/abs/1605.06966}{{\ttfamily
  arXiv:1605.06966 [nucl-ex]}}.

\bibitem{Eskola:2022vpi}
K.~J. Eskola, C.~A. Flett, V.~Guzey, T.~L\"oyt\"ainen, and H.~Paukkunen,
  ``{Exclusive $J/\psi$ photoproduction in ultraperipheral Pb+Pb collisions at
  the LHC to next-to-leading order perturbative QCD}'',
  \href{http://arxiv.org/abs/2203.11613}{{\ttfamily arXiv:2203.11613
  [hep-ph]}}.

\bibitem{Citron:2018lsq}
Z.~Citron {\em et~al.}, ``{Report from Working Group 5}: {Future physics
  opportunities for high-density QCD at the LHC with heavy-ion and proton
  beams}'', \href{http://dx.doi.org/10.23731/CYRM-2019-007.1159}{{\em CERN
  Yellow Rep. Monogr.} {\bfseries 7} (2019) },
  \href{http://arxiv.org/abs/1812.06772}{{\ttfamily arXiv:1812.06772
  [hep-ph]}}.

\bibitem{Hentschinski:2022xnd}
M.~Hentschinski {\em et~al.}, ``{White Paper on Forward Physics, BFKL,
  Saturation Physics and Diffraction}'',
  \href{http://arxiv.org/abs/2203.08129}{{\ttfamily arXiv:2203.08129
  [hep-ph]}}.

\bibitem{Klein:2002wm}
S.~R. Klein, J.~Nystrand, and R.~Vogt, ``{Heavy quark photoproduction in
  ultraperipheral heavy ion collisions}'',
  \href{http://dx.doi.org/10.1103/PhysRevC.66.044906}{{\em Phys. Rev. C}
  {\bfseries 66} (2002) },
  \href{http://arxiv.org/abs/hep-ph/0206220}{{\ttfamily arXiv:hep-ph/0206220}}.

\bibitem{ATLAS:2017kwa}
{\bfseries ATLAS} Collaboration, {ATLAS Collaboration}, ``{Photo-nuclear dijet
  production in ultra-peripheral Pb+Pb collisions}.'' {ATLAS-CONF-2017-011},
2017.

\bibitem{CMS:2022lbi}
{\bfseries CMS} Collaboration, ``{Azimuthal correlations within exclusive
  dijets with large momentum transfer in photon-lead collisions}'',
  \href{http://arxiv.org/abs/2205.00045}{{\ttfamily arXiv:2205.00045
  [nucl-ex]}}.

\bibitem{STAR:2017enh}
{\bfseries STAR} Collaboration, L.~Adamczyk {\em et~al.}, ``{Coherent
  diffractive photoproduction of $\rho^{0}$mesons on gold nuclei at 200
  GeV/nucleon-pair at the Relativistic Heavy Ion Collider}'',
  \href{http://dx.doi.org/10.1103/PhysRevC.96.054904}{{\em Phys. Rev. C}
  {\bfseries 96} no.~5, (2017) },
  \href{http://arxiv.org/abs/1702.07705}{{\ttfamily arXiv:1702.07705
  [nucl-ex]}}.

\bibitem{Klein:2021mgd}
{\bfseries STAR} Collaboration, S.~R. Klein, ``{Using coherent dipion
  photoproduction to image gold nuclei}'',
  \href{http://dx.doi.org/10.21468/SciPostPhysProc.8.128}{{\em SciPost Phys.
  Proc.} {\bfseries 8} (2022) },
  \href{http://arxiv.org/abs/2107.10447}{{\ttfamily arXiv:2107.10447
  [nucl-ex]}}.

\bibitem{Cepila:2016uku}
J.~Cepila, J.~G. Contreras, and J.~D. Tapia~Takaki, ``{Energy dependence of
  dissociative $\mathrm{J/}\psi$ photoproduction as a signature of gluon
  saturation at the LHC}'',
  \href{http://dx.doi.org/10.1016/j.physletb.2016.12.063}{{\em Phys. Lett. B}
  {\bfseries 766} (2017) }, \href{http://arxiv.org/abs/1608.07559}{{\ttfamily
  arXiv:1608.07559 [hep-ph]}}.

\bibitem{Cepila:2018zky}
J.~Cepila, J.~G. Contreras, M.~Krelina, and J.~D. Tapia~Takaki, ``{Mass
  dependence of vector meson photoproduction off protons and nuclei within the
  energy-dependent hot-spot model}'',
  \href{http://dx.doi.org/10.1016/j.nuclphysb.2018.07.010}{{\em Nucl. Phys. B}
  {\bfseries 934} (2018) }, \href{http://arxiv.org/abs/1804.05508}{{\ttfamily
  arXiv:1804.05508 [hep-ph]}}.

\bibitem{Klein:2019qfb}
S.~R. Klein and H.~M\"antysaari, ``{Imaging the nucleus with high-energy
  photons}'', \href{http://dx.doi.org/10.1038/s42254-019-0107-6}{{\em Nature
  Rev. Phys.} {\bfseries 1} no.~11, (2019) },
  \href{http://arxiv.org/abs/1910.10858}{{\ttfamily arXiv:1910.10858
  [hep-ex]}}.

\bibitem{Broz:2019kpl}
M.~Broz, J.~G. Contreras, and J.~D. Tapia~Takaki, ``{A generator of forward
  neutrons for ultra-peripheral collisions:
  $\textbf{n}^{\textbf{O}}_{\textbf{O}}\textbf{n}$}'',
  \href{http://dx.doi.org/10.1016/j.cpc.2020.107181}{{\em Comput. Phys.
  Commun.} {\bfseries 253} (2020) },
  \href{http://arxiv.org/abs/1908.08263}{{\ttfamily arXiv:1908.08263
  [nucl-th]}}.

\bibitem{Bylinkin:2022temp}
A.~Bylinkin, J.~Nystrand, and D.~Tapia~Takaki, ``{Vector meson photoproduction
  in UPCs with FoCal}'', \href{http://arxiv.org/abs/2211.16107}{{\ttfamily
  arXiv:2211.16107 [nucl-ex]}}.

\bibitem{Martens:2017cvj}
J.~C. Martens, J.~P. Ralston, and J.~D. Tapia~Takaki, ``{Quantum tomography for
  collider physics: Illustrations with lepton pair production}'',
  \href{http://dx.doi.org/10.1140/epjc/s10052-017-5455-8}{{\em Eur. Phys. J. C}
  {\bfseries 78} no.~1, (2018) },
  \href{http://arxiv.org/abs/1707.01638}{{\ttfamily arXiv:1707.01638
  [hep-ph]}}.

\bibitem{Klusek-Gawenda:2020gwa}
M.~Klusek-Gawenda and J.~D. Tapia~Takaki, ``{Exclusive Four-pion
  Photoproduction in Ultra-peripheral Heavy-ion Collisions at RHIC and LHC
  Energies}'', \href{http://dx.doi.org/10.5506/APhysPolB.51.1393}{{\em Acta
  Phys. Polon. B} {\bfseries 51} no.~6, (2020) },
  \href{http://arxiv.org/abs/2005.13624}{{\ttfamily arXiv:2005.13624
  [hep-ph]}}.

\bibitem{ALICE:2018lao}
{\bfseries ALICE} Collaboration, S.~Acharya {\em et~al.}, ``{Anisotropic flow
  in Xe-Xe collisions at $\mathbf{\sqrt{s_{\rm{NN}}} = 5.44}$ TeV}'',
  \href{http://dx.doi.org/10.1016/j.physletb.2018.06.059}{{\em Phys. Lett. B}
  {\bfseries 784} (2018) }, \href{http://arxiv.org/abs/1805.01832}{{\ttfamily
  arXiv:1805.01832 [nucl-ex]}}.

\bibitem{Giacalone:2021udy}
G.~Giacalone, J.~Jia, and C.~Zhang, ``{Impact of Nuclear Deformation on
  Relativistic Heavy-Ion Collisions: Assessing Consistency in Nuclear Physics
  across Energy Scales}'',
  \href{http://dx.doi.org/10.1103/PhysRevLett.127.242301}{{\em Phys. Rev.
  Lett.} {\bfseries 127} no.~24, (2021) },
  \href{http://arxiv.org/abs/2105.01638}{{\ttfamily arXiv:2105.01638
  [nucl-th]}}.

\bibitem{Summerfield:2021oex}
N.~Summerfield, B.-N. Lu, C.~Plumberg, D.~Lee, J.~Noronha-Hostler, and
  A.~Timmins, ``{$^{16}$O $^{16}$O collisions at energies available at the BNL
  Relativistic Heavy Ion Collider and at the CERN Large Hadron Collider
  comparing $\alpha$ clustering versus substructure}'',
  \href{http://dx.doi.org/10.1103/PhysRevC.104.L041901}{{\em Phys. Rev. C}
  {\bfseries 104} no.~4, (2021) },
  \href{http://arxiv.org/abs/2103.03345}{{\ttfamily arXiv:2103.03345
  [nucl-th]}}.

\bibitem{Jia:2022qgl}
J.~Jia, G.~Giacalone, and C.~Zhang, ``{Separating the impact of nuclear skin
  and nuclear deformation on elliptic flow and its fluctuations in high-energy
  isobar collisions}'', \href{http://arxiv.org/abs/2206.10449}{{\ttfamily
  arXiv:2206.10449 [nucl-th]}}.

\bibitem{Giannini:2022lbj}
A.~V. Giannini, M.~N. Ferreira, M.~Hippert, D.~D. Chinellato, G.~S. Denicol,
  M.~Luzum, J.~Noronha, T.~N. da~Silva, and J.~Takahashi, ``{Assessing the
  ultra-central flow puzzle in the Bayesian era}'',
  \href{http://arxiv.org/abs/2203.17011}{{\ttfamily arXiv:2203.17011
  [nucl-th]}}.

\bibitem{Carzon:2020xwp}
P.~Carzon, S.~Rao, M.~Luzum, M.~Sievert, and J.~Noronha-Hostler, ``{Possible
  octupole deformation of $^{208}$Pb and the ultracentral $v_2$ to $v_3$
  puzzle}'', \href{http://dx.doi.org/10.1103/PhysRevC.102.054905}{{\em Phys.
  Rev. C} {\bfseries 102} no.~5, (2020) },
  \href{http://arxiv.org/abs/2007.00780}{{\ttfamily arXiv:2007.00780
  [nucl-th]}}.

\bibitem{Raman:2001nnq}
S.~Raman, C.~W.~G. Nestor, Jr, and P.~Tikkanen, ``{Transition probability from
  the ground to the first-excited 2+ state of even-even nuclides}'',
  \href{http://dx.doi.org/10.1006/adnd.2001.0858}{{\em Atom. Data Nucl. Data
  Tabl.} {\bfseries 78} (2001) }.

\bibitem{ALICE:2014dwt}
{\bfseries ALICE} Collaboration, B.~B. Abelev {\em et~al.}, ``{Multiparticle
  azimuthal correlations in p -Pb and Pb-Pb collisions at the CERN Large Hadron
  Collider}'', \href{http://dx.doi.org/10.1103/PhysRevC.90.054901}{{\em Phys.
  Rev. C} {\bfseries 90} no.~5, (2014) },
  \href{http://arxiv.org/abs/1406.2474}{{\ttfamily arXiv:1406.2474 [nucl-ex]}}.

\bibitem{Finazzo:2014cna}
S.~I. Finazzo, R.~Rougemont, H.~Marrochio, and J.~Noronha, ``{Hydrodynamic
  transport coefficients for the non-conformal quark-gluon plasma from
  holography}'', \href{http://dx.doi.org/10.1007/JHEP02(2015)051}{{\em JHEP}
  {\bfseries 02} (2015) }, \href{http://arxiv.org/abs/1412.2968}{{\ttfamily
  arXiv:1412.2968 [hep-ph]}}.

\bibitem{Ratti:2018ksb}
C.~Ratti, ``{Lattice QCD and heavy ion collisions: a review of recent
  progress}'', \href{http://dx.doi.org/10.1088/1361-6633/aabb97}{{\em Rept.
  Prog. Phys.} {\bfseries 81} no.~8, (2018) },
  \href{http://arxiv.org/abs/1804.07810}{{\ttfamily arXiv:1804.07810
  [hep-lat]}}.

\bibitem{Floerchinger:2015efa}
S.~Floerchinger and M.~Martinez, ``{Fluid dynamic propagation of initial baryon
  number perturbations on a Bjorken flow background}'',
  \href{http://dx.doi.org/10.1103/PhysRevC.92.064906}{{\em Phys. Rev. C}
  {\bfseries 92} no.~6, (2015) },
  \href{http://arxiv.org/abs/1507.05569}{{\ttfamily arXiv:1507.05569
  [nucl-th]}}.

\bibitem{ALICE:2021hjb}
{\bfseries ALICE} Collaboration, S.~Acharya {\em et~al.}, ``{General balance
  functions of identified charged hadron pairs of ($pi$,K,p) in Pb-Pb
  collisions at $\sqrt{s_{\rm NN}} =2.76$ TeV}'',
  \href{http://dx.doi.org/10.1016/j.physletb.2022.137338}{{\em Phys. Lett. B}
  {\bfseries 833} (2022) }, \href{http://arxiv.org/abs/2110.06566}{{\ttfamily
  arXiv:2110.06566 [nucl-ex]}}.

\bibitem{Pratt:2021xvg}
S.~Pratt and C.~Plumberg, ``{Charge balance functions for heavy-ion collisions
  at energies available at the CERN Large Hadron Collider}'',
  \href{http://dx.doi.org/10.1103/PhysRevC.104.014906}{{\em Phys. Rev. C}
  {\bfseries 104} no.~1, (2021) },
  \href{http://arxiv.org/abs/2104.00628}{{\ttfamily arXiv:2104.00628
  [nucl-th]}}.

\bibitem{ALICE:2010yje}
{\bfseries ALICE} Collaboration, K.~Aamodt {\em et~al.}, ``{Suppression of
  Charged Particle Production at Large Transverse Momentum in Central Pb-Pb
  Collisions at $\sqrt{s_{NN}} =$ 2.76 TeV}'',
  \href{http://dx.doi.org/10.1016/j.physletb.2010.12.020}{{\em Phys. Lett. B}
  {\bfseries 696} (2011) }, \href{http://arxiv.org/abs/1012.1004}{{\ttfamily
  arXiv:1012.1004 [nucl-ex]}}.

\bibitem{ALICE:2018vuu}
{\bfseries ALICE} Collaboration, S.~Acharya {\em et~al.}, ``{Transverse
  momentum spectra and nuclear modification factors of charged particles in pp,
  p-Pb and Pb-Pb collisions at the LHC}'',
  \href{http://dx.doi.org/10.1007/JHEP11(2018)013}{{\em JHEP} {\bfseries 11}
  (2018) }, \href{http://arxiv.org/abs/1802.09145}{{\ttfamily arXiv:1802.09145
  [nucl-ex]}}.

\bibitem{ALICE:2019qyj}
{\bfseries ALICE} Collaboration, S.~Acharya {\em et~al.}, ``{Measurements of
  inclusive jet spectra in pp and central Pb-Pb collisions at
  $\sqrt{s_{\rm{NN}}}$ = 5.02 TeV}'',
  \href{http://dx.doi.org/10.1103/PhysRevC.101.034911}{{\em Phys. Rev. C}
  {\bfseries 101} no.~3, (2020) },
  \href{http://arxiv.org/abs/1909.09718}{{\ttfamily arXiv:1909.09718
  [nucl-ex]}}.

\bibitem{RCTaliceQM22}
{Reynier Cruz Torres for the ALICE Collaboration}, ``Jet acoplanarity and
  energy flow within jets in pb-pb and pp collisions with alice.''
  \textsc{url:}~\url{https://indico.cern.ch/event/895086/contributions/4742231/},
  4, 2022.
\newblock Quark Matter 2022 Conference.

\bibitem{HBaliceQM22}
{Hannah Bossi for the ALICE Collaboration}, ``Measurements of the
  $r$-dependence of inclusive jet suppression and groomed jet splittings in
  heavy-ion collisions with alice.''
  \textsc{url:}~\url{https://indico.cern.ch/event/895086/contributions/4736731/},
  4, 2022.
\newblock Quark Matter 2022 Conference.

\bibitem{Haake:2018hqn}
R.~Haake and C.~Loizides, ``{Machine Learning based jet momentum reconstruction
  in heavy-ion collisions}'',
  \href{http://dx.doi.org/10.1103/PhysRevC.99.064904}{{\em Phys. Rev. C}
  {\bfseries 99} no.~6, (2019) },
  \href{http://arxiv.org/abs/1810.06324}{{\ttfamily arXiv:1810.06324
  [nucl-ex]}}.

\bibitem{ALICE:2015mdb}
{\bfseries ALICE} Collaboration, J.~Adam {\em et~al.}, ``{Measurement of jet
  quenching with semi-inclusive hadron-jet distributions in central Pb-Pb
  collisions at $ \sqrt{s_{\mathrm{NN}}}=2.76 $ TeV}'',
  \href{http://dx.doi.org/10.1007/JHEP09(2015)170}{{\em JHEP} {\bfseries 09}
  (2015) }, \href{http://arxiv.org/abs/1506.03984}{{\ttfamily arXiv:1506.03984
  [nucl-ex]}}.

\bibitem{Cao:2020wlm}
S.~Cao and X.-N. Wang, ``{Jet quenching and medium response in high-energy
  heavy-ion collisions: a review}'',
  \href{http://dx.doi.org/10.1088/1361-6633/abc22b}{{\em Rept. Prog. Phys.}
  {\bfseries 84} no.~2, (2021) },
  \href{http://arxiv.org/abs/2002.04028}{{\ttfamily arXiv:2002.04028
  [hep-ph]}}.

\bibitem{DEramo:2018eoy}
F.~D'Eramo, K.~Rajagopal, and Y.~Yin, ``{Moli\`ere scattering in quark-gluon
  plasma: finding point-like scatterers in a liquid}'',
  \href{http://dx.doi.org/10.1007/JHEP01(2019)172}{{\em JHEP} {\bfseries 01}
  (2019) }, \href{http://arxiv.org/abs/1808.03250}{{\ttfamily arXiv:1808.03250
  [hep-ph]}}.

\bibitem{Barata:2020rdn}
J.~a. Barata, Y.~Mehtar-Tani, A.~Soto-Ontoso, and K.~Tywoniuk, ``{Revisiting
  transverse momentum broadening in dense QCD media}'',
  \href{http://dx.doi.org/10.1103/PhysRevD.104.054047}{{\em Phys. Rev. D}
  {\bfseries 104} no.~5, (2021) },
  \href{http://arxiv.org/abs/2009.13667}{{\ttfamily arXiv:2009.13667
  [hep-ph]}}.

\bibitem{Barata:2021wuf}
J.~a. Barata, Y.~Mehtar-Tani, A.~Soto-Ontoso, and K.~Tywoniuk,
  ``{Medium-induced radiative kernel with the Improved Opacity Expansion}'',
  \href{http://dx.doi.org/10.1007/JHEP09(2021)153}{{\em JHEP} {\bfseries 09}
  (2021) }, \href{http://arxiv.org/abs/2106.07402}{{\ttfamily arXiv:2106.07402
  [hep-ph]}}.

\bibitem{ALargeIonColliderExperiment:2021mqf}
{\bfseries A Large Ion Collider Experiment, ALICE} Collaboration, S.~Acharya
  {\em et~al.}, ``{Measurement of the groomed jet radius and momentum splitting
  fraction in pp and Pb$-$Pb collisions at $\sqrt{s_{NN}} = 5.02$ TeV}'',
  \href{http://dx.doi.org/10.1103/PhysRevLett.128.102001}{{\em Phys. Rev.
  Lett.} {\bfseries 128} no.~10, (2022) },
  \href{http://arxiv.org/abs/2107.12984}{{\ttfamily arXiv:2107.12984
  [nucl-ex]}}.

\bibitem{ALICE:2022vsz}
{\bfseries ALICE} Collaboration, ``{Measurement of inclusive and leading subjet
  fragmentation in pp and Pb-Pb collisions at $\sqrt{s_{\rm NN}}$ = 5.02
  TeV}'', \href{http://arxiv.org/abs/2204.10270}{{\ttfamily arXiv:2204.10270
  [nucl-ex]}}.

\bibitem{JMaliceQM22}
{James Mulligan for the ALICE Collaboration}, ``Jet angularity and
  fragmentation measurements in heavy-ion collisions with alice.''
  \textsc{url:}~\url{https://indico.cern.ch/event/895086/contributions/4715856/},
  4, 2022.
\newblock Quark Matter 2022 Conference.

\bibitem{ALICE:2022hyz}
{\bfseries ALICE} Collaboration, ``{Measurements of the groomed jet radius and
  momentum splitting fraction with the soft drop and dynamical grooming
  algorithms in pp collisions at $\sqrt{s}=5.02$ TeV}'',
  \href{http://arxiv.org/abs/2204.10246}{{\ttfamily arXiv:2204.10246
  [nucl-ex]}}.

\bibitem{ALICE:2021yet}
{\bfseries ALICE} Collaboration, ``{Physics Preliminary Summary: Measurement of
  the primary Lund plane density in pp collisions at $\sqrt{s} = \rm{13}$ TeV
  with ALICE}'',.

\bibitem{Ringer:2019rfk}
F.~Ringer, B.-W. Xiao, and F.~Yuan, ``{Can we observe jet $P_T$-broadening in
  heavy-ion collisions at the LHC?}'',
  \href{http://dx.doi.org/10.1016/j.physletb.2020.135634}{{\em Phys. Lett. B}
  {\bfseries 808} (2020) }, \href{http://arxiv.org/abs/1907.12541}{{\ttfamily
  arXiv:1907.12541 [hep-ph]}}.

\bibitem{Lifson:2020gua}
A.~Lifson, G.~P. Salam, and G.~Soyez, ``{Calculating the primary Lund Jet Plane
  density}'', \href{http://dx.doi.org/10.1007/JHEP10(2020)170}{{\em JHEP}
  {\bfseries 10} (2020) }, \href{http://arxiv.org/abs/2007.06578}{{\ttfamily
  arXiv:2007.06578 [hep-ph]}}.

\bibitem{ALICE:2019sqi}
{\bfseries ALICE} Collaboration, S.~Acharya {\em et~al.}, ``{Jet-hadron
  correlations measured relative to the second order event plane in Pb-Pb
  collisions at $\sqrt{s_{\rm{NN}}}$ = 2.76 TeV}'',
  \href{http://dx.doi.org/10.1103/PhysRevC.101.064901}{{\em Phys. Rev. C}
  {\bfseries 101} no.~6, (2020) },
  \href{http://arxiv.org/abs/1910.14398}{{\ttfamily arXiv:1910.14398
  [nucl-ex]}}.

\bibitem{ALICE:2015efi}
{\bfseries ALICE} Collaboration, J.~Adam {\em et~al.}, ``{Azimuthal anisotropy
  of charged jet production in $\sqrt{s_{\rm NN}}$ = 2.76 TeV Pb-Pb
  collisions}'', \href{http://dx.doi.org/10.1016/j.physletb.2015.12.047}{{\em
  Phys. Lett. B} {\bfseries 753} (2016) },
  \href{http://arxiv.org/abs/1509.07334}{{\ttfamily arXiv:1509.07334
  [nucl-ex]}}.

\bibitem{Schukraft:2012ah}
J.~Schukraft, A.~Timmins, and S.~A. Voloshin, ``{Ultra-relativistic nuclear
  collisions: event shape engineering}'',
  \href{http://dx.doi.org/10.1016/j.physletb.2013.01.045}{{\em Phys. Lett. B}
  {\bfseries 719} (2013) }, \href{http://arxiv.org/abs/1208.4563}{{\ttfamily
  arXiv:1208.4563 [nucl-ex]}}.

\bibitem{CBaliceQM22}
{Caitlin Beattie for the ALICE Collaboration}, ``Study of path-length dependent
  energy loss of jets in p-pb and pb-pb collisions with alice.''
  \textsc{url:}~\url{https://indico.cern.ch/event/895086/contributions/4715857/},
  4, 2022.
\newblock Quark Matter 2022 Conference.

\bibitem{ALaliceQM22}
{Alwina Liu for the ALICE Collaboration}, ``Isolated-photon production and
  photon-jet correlations in pp and pb-pb collisions at $\sqrt{s_{\rm NN}} =$
  5.02 tev in alice.''
  \textsc{url:}~\url{https://indico.cern.ch/event/895086/contributions/4715853/},
  4, 2022.
\newblock Quark Matter 2022 Conference.

\bibitem{ALICE:2021rxa}
{\bfseries ALICE} Collaboration, S.~Acharya {\em et~al.}, ``{Prompt D$^{0}$,
  D$^{+}$, and D$^{*+}$ production in Pb\textendash{}Pb collisions at $
  \sqrt{s_{\mathrm{NN}}} $ = 5.02 TeV}'',
  \href{http://dx.doi.org/10.1007/JHEP01(2022)174}{{\em JHEP} {\bfseries 01}
  (2022) }, \href{http://arxiv.org/abs/2110.09420}{{\ttfamily arXiv:2110.09420
  [nucl-ex]}}.

\bibitem{ALICE:2022tji}
{\bfseries ALICE} Collaboration, S.~Acharya {\em et~al.}, ``{Measurement of
  beauty production via non-prompt ${\rm D}^{0}$ mesons in Pb-Pb collisions at
  $\sqrt{s_{\rm NN}}$ = 5.02 TeV}'',
  \href{http://arxiv.org/abs/2202.00815}{{\ttfamily arXiv:2202.00815
  [nucl-ex]}}.

\bibitem{ALICE:2019nuy}
{\bfseries ALICE} Collaboration, S.~Acharya {\em et~al.}, ``{Measurement of
  electrons from semileptonic heavy-flavour hadron decays at midrapidity in pp
  and Pb-Pb collisions at $\sqrt{s_{\rm{NN}}}$ = 5.02 TeV}'',
  \href{http://dx.doi.org/10.1016/j.physletb.2020.135377}{{\em Phys. Lett. B}
  {\bfseries 804} (2020) }, \href{http://arxiv.org/abs/1910.09110}{{\ttfamily
  arXiv:1910.09110 [nucl-ex]}}.

\bibitem{Voutilainen:2015lqa}
M.~Voutilainen, ``{Heavy quark jets at the LHC}'',
  \href{http://dx.doi.org/10.1142/S0217751X15460082}{{\em Int. J. Mod. Phys. A}
  {\bfseries 30} no.~31, (2015) },
  \href{http://arxiv.org/abs/1509.05026}{{\ttfamily arXiv:1509.05026
  [hep-ex]}}.

\bibitem{JETSCAPE:2021ehl}
{\bfseries JETSCAPE} Collaboration, S.~Cao {\em et~al.}, ``{Determining the jet
  transport coefficient from inclusive hadron suppression measurements using
  Bayesian parameter estimation}'',
  \href{http://dx.doi.org/10.1103/PhysRevC.104.024905}{{\em Phys. Rev. C}
  {\bfseries 104} no.~2, (2021) },
  \href{http://arxiv.org/abs/2102.11337}{{\ttfamily arXiv:2102.11337
  [nucl-th]}}.

\bibitem{JETSCAPE:2022ixz}
{\bfseries JETSCAPE} Collaboration, R.~Ehlers {\em et~al.}, ``{Bayesian
  analysis of QGP jet transport using multi-scale modeling applied to inclusive
  hadron and reconstructed jet data}'', in {\em {29th International Conference
  on Ultra-relativistic Nucleus-Nucleus Collisions}}.
\newblock 8, 2022.
\newblock \href{http://arxiv.org/abs/2208.07950}{{\ttfamily arXiv:2208.07950
  [hep-ph]}}.

\bibitem{Cao:2017crw}
S.~Cao, A.~Majumder, G.-Y. Qin, and C.~Shen, ``{Drag Induced Radiation and
  Multi-Stage Effects in Heavy-Flavor Energy Loss}'',
  \href{http://dx.doi.org/10.1016/j.physletb.2019.05.020}{{\em Phys. Lett. B}
  {\bfseries 793} (2019) }, \href{http://arxiv.org/abs/1711.09053}{{\ttfamily
  arXiv:1711.09053 [nucl-th]}}.

\bibitem{Moore:2004tg}
G.~D. Moore and D.~Teaney, ``{How much do heavy quarks thermalize in a heavy
  ion collision?}'', \href{http://dx.doi.org/10.1103/PhysRevC.71.064904}{{\em
  Phys. Rev. C} {\bfseries 71} (2005) },
  \href{http://arxiv.org/abs/hep-ph/0412346}{{\ttfamily arXiv:hep-ph/0412346}}.

\bibitem{Dreyer:2021hhr}
F.~A. Dreyer, G.~Soyez, and A.~Takacs, ``{Quarks and gluons in the Lund
  plane}'', \href{http://dx.doi.org/10.1007/JHEP08(2022)177}{{\em JHEP}
  {\bfseries 08} (2022) }, \href{http://arxiv.org/abs/2112.09140}{{\ttfamily
  arXiv:2112.09140 [hep-ph]}}.

\bibitem{ATLAS:2022zgo}
{\bfseries ATLAS} Collaboration, ``{Dependence of the Jet Energy Scale on the
  Particle Content of Hadronic Jets in the ATLAS Detector Simulation}'',.

\bibitem{Sapeta:2007ad}
S.~Sapeta and U.~A. Wiedemann, ``{Jet hadrochemistry as a characteristics of
  jet quenching}'',
  \href{http://dx.doi.org/10.1140/epjc/s10052-008-0592-8}{{\em Eur. Phys. J. C}
  {\bfseries 55} (2008) }, \href{http://arxiv.org/abs/0707.3494}{{\ttfamily
  arXiv:0707.3494 [hep-ph]}}.

\bibitem{Attems:2022ubu}
M.~Attems, J.~Brewer, G.~M. Innocenti, A.~Mazeliauskas, S.~Park, W.~van~der
  Schee, and U.~A. Wiedemann, ``{The medium-modified $g\to c\bar{c}$ splitting
  function in the BDMPS-Z formalism}'',
  \href{http://arxiv.org/abs/2203.11241}{{\ttfamily arXiv:2203.11241
  [hep-ph]}}.

\bibitem{Lai:2021ckt}
Y.~S. Lai, J.~Mulligan, M.~P\l{}osko\'n, and F.~Ringer, ``{The information
  content of jet quenching and machine learning assisted observable design}'',
  \href{http://dx.doi.org/10.1007/JHEP10(2022)011}{{\em JHEP} {\bfseries 10}
  (2022) }, \href{http://arxiv.org/abs/2111.14589}{{\ttfamily arXiv:2111.14589
  [hep-ph]}}.

\bibitem{Andreassen:2019cjw}
A.~Andreassen, P.~T. Komiske, E.~M. Metodiev, B.~Nachman, and J.~Thaler,
  ``{OmniFold: A Method to Simultaneously Unfold All Observables}'',
  \href{http://dx.doi.org/10.1103/PhysRevLett.124.182001}{{\em Phys. Rev.
  Lett.} {\bfseries 124} no.~18, (2020) },
  \href{http://arxiv.org/abs/1911.09107}{{\ttfamily arXiv:1911.09107
  [hep-ph]}}.

\bibitem{Mulligan:2020tim}
J.~Mulligan and M.~Ploskon, ``{Identifying groomed jet splittings in heavy-ion
  collisions}'', \href{http://dx.doi.org/10.1103/PhysRevC.102.044913}{{\em
  Phys. Rev. C} {\bfseries 102} no.~4, (2020) },
  \href{http://arxiv.org/abs/2006.01812}{{\ttfamily arXiv:2006.01812
  [hep-ph]}}.

\bibitem{Nijs:2020roc}
G.~Nijs, W.~van~der Schee, U.~G\"ursoy, and R.~Snellings, ``{Bayesian analysis
  of heavy ion collisions with the heavy ion computational framework
  Trajectum}'', \href{http://dx.doi.org/10.1103/PhysRevC.103.054909}{{\em Phys.
  Rev. C} {\bfseries 103} no.~5, (2021) },
  \href{http://arxiv.org/abs/2010.15134}{{\ttfamily arXiv:2010.15134
  [nucl-th]}}.

\bibitem{Chien:2018dfn}
Y.-T. Chien and R.~Kunnawalkam~Elayavalli, ``{Probing heavy ion collisions
  using quark and gluon jet substructure}'',
  \href{http://arxiv.org/abs/1803.03589}{{\ttfamily arXiv:1803.03589
  [hep-ph]}}.

\bibitem{Apolinario:2021olp}
L.~Apolin\'ario, N.~F. Castro, M.~Crispim Rom\~ao, J.~G. Milhano, R.~Pedro, and
  F.~C.~R. Peres, ``{Deep Learning for the classification of quenched jets}'',
  \href{http://dx.doi.org/10.1007/JHEP11(2021)219}{{\em JHEP} {\bfseries 11}
  (2021) }, \href{http://arxiv.org/abs/2106.08869}{{\ttfamily arXiv:2106.08869
  [hep-ph]}}.

\bibitem{Liu:2022hzd}
L.~Liu, J.~Velkovska, and M.~Verweij, ``{Identifying quenched jets in heavy ion
  collisions with machine learning}'',
  \href{http://arxiv.org/abs/2206.01628}{{\ttfamily arXiv:2206.01628
  [hep-ph]}}.

\bibitem{Andersson:1983ia}
B.~Andersson, G.~Gustafson, G.~Ingelman, and T.~Sjostrand, ``{Parton
  Fragmentation and String Dynamics}'',
  \href{http://dx.doi.org/10.1016/0370-1573(83)90080-7}{{\em Phys. Rept.}
  {\bfseries 97} (1983) }.

\bibitem{ALICE:2020wfu}
{\bfseries ALICE} Collaboration, S.~Acharya {\em et~al.}, ``{$\Lambda^+_c$
  Production and Baryon-to-Meson Ratios in pp and p-Pb Collisions at $\sqrt
  {s_{NN}}$=5.02\,\,TeV at the LHC}'',
  \href{http://dx.doi.org/10.1103/PhysRevLett.127.202301}{{\em Phys. Rev.
  Lett.} {\bfseries 127} no.~20, (2021) },
  \href{http://arxiv.org/abs/2011.06078}{{\ttfamily arXiv:2011.06078
  [nucl-ex]}}.

\bibitem{Tu:2019ouv}
Z.~Tu, D.~E. Kharzeev, and T.~Ullrich, ``{Einstein-Podolsky-Rosen Paradox and
  Quantum Entanglement at Subnucleonic Scales}'',
  \href{http://dx.doi.org/10.1103/PhysRevLett.124.062001}{{\em Phys. Rev.
  Lett.} {\bfseries 124} no.~6, (2020) },
  \href{http://arxiv.org/abs/1904.11974}{{\ttfamily arXiv:1904.11974
  [hep-ph]}}.

\bibitem{Flor:2020fdw}
F.~A. Flor, G.~Olinger, and R.~Bellwied, ``{Flavour and Energy Dependence of
  Chemical Freeze-out Temperatures in Relativistic Heavy Ion Collisions from
  RHIC-BES to LHC Energies}'',
  \href{http://dx.doi.org/10.1016/j.physletb.2021.136098}{{\em Phys. Lett. B}
  {\bfseries 814} (2021) }, \href{http://arxiv.org/abs/2009.14781}{{\ttfamily
  arXiv:2009.14781 [nucl-ex]}}.

\bibitem{ALICE:2021bib}
{\bfseries ALICE} Collaboration, S.~Acharya {\em et~al.}, ``{Constraining
  hadronization mechanisms with $\rm \Lambda_{\rm c}^{+}$/D$^0$ production
  ratios in Pb-Pb collisions at $\sqrt{s_{\rm NN}} = 5.02$ TeV}'',
  \href{http://arxiv.org/abs/2112.08156}{{\ttfamily arXiv:2112.08156
  [nucl-ex]}}.

\bibitem{ALICE:2021vxl}
{\bfseries ALICE} Collaboration, S.~Acharya {\em et~al.}, ``{Production of
  $\Lambda$ and $K^0_s$ in jets in p\textendash{}Pb collisions at $\sqrt
  {s_{NN}}$=5.02 TeV and pp collisions at $\sqrt {s}$=7 TeV}'',
  \href{http://dx.doi.org/10.1016/j.physletb.2022.136984}{{\em Phys. Lett. B}
  {\bfseries 827} (2022) }, \href{http://arxiv.org/abs/2105.04890}{{\ttfamily
  arXiv:2105.04890 [nucl-ex]}}.

\bibitem{Andronic:2017pug}
A.~Andronic, P.~Braun-Munzinger, K.~Redlich, and J.~Stachel, ``{Decoding the
  phase structure of QCD via particle production at high energy}'',
  \href{http://dx.doi.org/10.1038/s41586-018-0491-6}{{\em Nature} {\bfseries
  561} no.~7723, (2018) }, \href{http://arxiv.org/abs/1710.09425}{{\ttfamily
  arXiv:1710.09425 [nucl-th]}}.

\bibitem{Jaffe:1976ig}
R.~L. Jaffe, ``{Multi-Quark Hadrons. 1. The Phenomenology of (2 Quark 2
  anti-Quark) Mesons}'', \href{http://dx.doi.org/10.1103/PhysRevD.15.267}{{\em
  Phys. Rev. D} {\bfseries 15} (1977) }.

\bibitem{ALICE:2017jto}
{\bfseries ALICE} Collaboration, S.~Acharya {\em et~al.}, ``{Measuring
  K$^0_{\rm S}$K$^{\rm \pm}$ interactions using Pb-Pb collisions at
  ${\sqrt{s_{\rm NN}}=2.76}$ TeV}'',
  \href{http://dx.doi.org/10.1016/j.physletb.2017.09.009}{{\em Phys. Lett. B}
  {\bfseries 774} (2017) }, \href{http://arxiv.org/abs/1705.04929}{{\ttfamily
  arXiv:1705.04929 [nucl-ex]}}.

\bibitem{ALICE:2018nnl}
{\bfseries ALICE} Collaboration, S.~Acharya {\em et~al.}, ``{Measuring
  K$^0_{\rm S}$K$^{\rm{\pm}}$ interactions using pp collisions at $\sqrt{s}=7$
  TeV}'', \href{http://dx.doi.org/10.1016/j.physletb.2018.12.033}{{\em Phys.
  Lett. B} {\bfseries 790} (2019) },
  \href{http://arxiv.org/abs/1809.07899}{{\ttfamily arXiv:1809.07899
  [nucl-ex]}}.

\bibitem{ALICE:2021ovd}
{\bfseries ALICE} Collaboration, S.~Acharya {\em et~al.}, ``{Measuring
  K$^0_{\rm S}$K$^{\rm{\pm}}$ interactions using pp collisions at $\sqrt{s}=5$
  and $\sqrt{s}=13$ TeV}'',
  \href{http://dx.doi.org/10.1016/j.physletb.2022.137335}{{\em Phys. Lett. B}
  {\bfseries 833} (2022) }, \href{http://arxiv.org/abs/2111.06611}{{\ttfamily
  arXiv:2111.06611 [nucl-ex]}}.

\bibitem{LHCb:2021vvq}
{\bfseries LHCb} Collaboration, R.~Aaij {\em et~al.}, ``{Observation of an
  exotic narrow doubly charmed tetraquark}'',
  \href{http://dx.doi.org/10.1038/s41567-022-01614-y}{{\em Nature Phys.}
  {\bfseries 18} no.~7, (2022) },
  \href{http://arxiv.org/abs/2109.01038}{{\ttfamily arXiv:2109.01038
  [hep-ex]}}.

\bibitem{LHCb:2021auc}
{\bfseries LHCb} Collaboration, R.~Aaij {\em et~al.}, ``{Study of the doubly
  charmed tetraquark $T_{cc}^{+}$}'',
  \href{http://dx.doi.org/10.1038/s41467-022-30206-w}{{\em Nature Commun.}
  {\bfseries 13} no.~1, (2022) },
  \href{http://arxiv.org/abs/2109.01056}{{\ttfamily arXiv:2109.01056
  [hep-ex]}}.

\bibitem{Andronic:2021erx}
A.~Andronic, P.~Braun-Munzinger, M.~K. K\"ohler, A.~Mazeliauskas, K.~Redlich,
  J.~Stachel, and V.~Vislavicius, ``{The multiple-charm hierarchy in the
  statistical hadronization model}'',
  \href{http://dx.doi.org/10.1007/JHEP07(2021)035}{{\em JHEP} {\bfseries 07}
  (2021) }, \href{http://arxiv.org/abs/2104.12754}{{\ttfamily arXiv:2104.12754
  [hep-ph]}}.

\bibitem{LHCb:2017iph}
{\bfseries LHCb} Collaboration, R.~Aaij {\em et~al.}, ``{Observation of the
  doubly charmed baryon $\Xi_{cc}^{++}$}'',
  \href{http://dx.doi.org/10.1103/PhysRevLett.119.112001}{{\em Phys. Rev.
  Lett.} {\bfseries 119} no.~11, (2017) },
  \href{http://arxiv.org/abs/1707.01621}{{\ttfamily arXiv:1707.01621
  [hep-ex]}}.

\bibitem{Hernandez:2019eox}
E.~Hern\'andez, J.~Vijande, A.~Valcarce, and J.-M. Richard, ``{Spectroscopy,
  lifetime and decay modes of the $T^-_{bb}$ tetraquark}'',
  \href{http://dx.doi.org/10.1016/j.physletb.2019.135073}{{\em Phys. Lett. B}
  {\bfseries 800} (2020) }, \href{http://arxiv.org/abs/1910.13394}{{\ttfamily
  arXiv:1910.13394 [hep-ph]}}.

\bibitem{Liu:2021gva}
Z.~Liu, H.-T. An, Z.-W. Liu, and X.~Liu, ``{Where are the hidden-charm
  hexaquarks?}'', \href{http://dx.doi.org/10.1103/PhysRevD.105.034006}{{\em
  Phys. Rev. D} {\bfseries 105} no.~3, (2022) },
  \href{http://arxiv.org/abs/2112.02510}{{\ttfamily arXiv:2112.02510
  [hep-ph]}}.

\bibitem{CMS:2021znk}
{\bfseries CMS} Collaboration, A.~M. Sirunyan {\em et~al.}, ``{Evidence for
  X(3872) in Pb-Pb Collisions and Studies of its Prompt Production at $\sqrt
  {s_{NN}}$=5.02\,\,TeV}'',
  \href{http://dx.doi.org/10.1103/PhysRevLett.128.032001}{{\em Phys. Rev.
  Lett.} {\bfseries 128} no.~3, (2022) },
  \href{http://arxiv.org/abs/2102.13048}{{\ttfamily arXiv:2102.13048
  [hep-ex]}}.

\bibitem{Andronic:2019wva}
A.~Andronic, P.~Braun-Munzinger, M.~K. K\"ohler, K.~Redlich, and J.~Stachel,
  ``{Transverse momentum distributions of charmonium states with the
  statistical hadronization model}'',
  \href{http://dx.doi.org/10.1016/j.physletb.2019.134836}{{\em Phys. Lett. B}
  {\bfseries 797} (2019) }, \href{http://arxiv.org/abs/1901.09200}{{\ttfamily
  arXiv:1901.09200 [nucl-th]}}.

\bibitem{ALICE:2020mfd}
{\bfseries ALICE} Collaboration, A.~Collaboration {\em et~al.}, ``{Unveiling
  the strong interaction among hadrons at the LHC}'',
  \href{http://dx.doi.org/10.1038/s41586-020-3001-6}{{\em Nature} {\bfseries
  588} (2020) }, \href{http://arxiv.org/abs/2005.11495}{{\ttfamily
  arXiv:2005.11495 [nucl-ex]}}. [Erratum: Nature 590, E13 (2021)].

\bibitem{ALICE:2012eyl}
{\bfseries ALICE} Collaboration, B.~Abelev {\em et~al.}, ``{Long-range angular
  correlations on the near and away side in $p$-Pb collisions at
  $\sqrt{s_{NN}}=5.02$ TeV}'',
  \href{http://dx.doi.org/10.1016/j.physletb.2013.01.012}{{\em Phys. Lett. B}
  {\bfseries 719} (2013) }, \href{http://arxiv.org/abs/1212.2001}{{\ttfamily
  arXiv:1212.2001 [nucl-ex]}}.

\bibitem{ALICE:2015lpx}
{\bfseries ALICE} Collaboration, J.~Adam {\em et~al.}, ``{Forward-central
  two-particle correlations in p-Pb collisions at $\sqrt{s_{\rm NN}}$ = 5.02
  TeV}'', \href{http://dx.doi.org/10.1016/j.physletb.2015.12.010}{{\em Phys.
  Lett. B} {\bfseries 753} (2016) },
  \href{http://arxiv.org/abs/1506.08032}{{\ttfamily arXiv:1506.08032
  [nucl-ex]}}.

\bibitem{ALICE:2017svf}
{\bfseries ALICE} Collaboration, S.~Acharya {\em et~al.}, ``{Constraints on jet
  quenching in p-Pb collisions at $\mathbf{\sqrt{s_{NN}}}$ = 5.02 TeV measured
  by the event-activity dependence of semi-inclusive hadron-jet
  distributions}'',
  \href{http://dx.doi.org/10.1016/j.physletb.2018.05.059}{{\em Phys. Lett. B}
  {\bfseries 783} (2018) }, \href{http://arxiv.org/abs/1712.05603}{{\ttfamily
  arXiv:1712.05603 [nucl-ex]}}.

\bibitem{Lee:2022kdn}
K.~Lee, J.~Mulligan, M.~P\l{}osko\'n, F.~Ringer, and F.~Yuan, ``{Machine
  learning-based jet and event classification at the Electron-Ion Collider with
  applications to hadron structure and spin physics}'',
  \href{http://arxiv.org/abs/2210.06450}{{\ttfamily arXiv:2210.06450
  [hep-ph]}}.

\bibitem{PHENIX:2018lia}
{\bfseries PHENIX} Collaboration, C.~Aidala {\em et~al.}, ``{Creation of
  quark\textendash{}gluon plasma droplets with three distinct geometries}'',
  \href{http://dx.doi.org/10.1038/s41567-018-0360-0}{{\em Nature Phys.}
  {\bfseries 15} no.~3, (2019) },
  \href{http://arxiv.org/abs/1805.02973}{{\ttfamily arXiv:1805.02973
  [nucl-ex]}}.

\bibitem{Schenke:2020mbo}
B.~Schenke, C.~Shen, and P.~Tribedy, ``{Running the gamut of high energy
  nuclear collisions}'',
  \href{http://dx.doi.org/10.1103/PhysRevC.102.044905}{{\em Phys. Rev. C}
  {\bfseries 102} no.~4, (2020) },
  \href{http://arxiv.org/abs/2005.14682}{{\ttfamily arXiv:2005.14682
  [nucl-th]}}.

\bibitem{Lim:2019cys}
S.~H. Lim, Q.~Hu, R.~Belmont, K.~K. Hill, J.~L. Nagle, and D.~V. Perepelitsa,
  ``{Examination of flow and nonflow factorization methods in small collision
  systems}'', \href{http://dx.doi.org/10.1103/PhysRevC.100.024908}{{\em Phys.
  Rev. C} {\bfseries 100} no.~2, (2019) },
  \href{http://arxiv.org/abs/1902.11290}{{\ttfamily arXiv:1902.11290
  [nucl-th]}}.

\bibitem{Voloshin:2008dg}
S.~A. Voloshin, A.~M. Poskanzer, and R.~Snellings, ``{Collective phenomena in
  non-central nuclear collisions}'',
  \href{http://dx.doi.org/10.1007/978-3-642-01539-7_10}{{\em Landolt-Bornstein}
  {\bfseries 23} (2010) }, \href{http://arxiv.org/abs/0809.2949}{{\ttfamily
  arXiv:0809.2949 [nucl-ex]}}.

\bibitem{Nagle:2018nvi}
J.~L. Nagle and W.~A. Zajc, ``{Small System Collectivity in Relativistic
  Hadronic and Nuclear Collisions}'',
  \href{http://dx.doi.org/10.1146/annurev-nucl-101916-123209}{{\em Ann. Rev.
  Nucl. Part. Sci.} {\bfseries 68} (2018) },
  \href{http://arxiv.org/abs/1801.03477}{{\ttfamily arXiv:1801.03477
  [nucl-ex]}}.

\bibitem{Bierlich:2015rha}
C.~Bierlich and J.~R. Christiansen, ``{Effects of color reconnection on hadron
  flavor observables}'',
  \href{http://dx.doi.org/10.1103/PhysRevD.92.094010}{{\em Phys. Rev. D}
  {\bfseries 92} no.~9, (2015) },
  \href{http://arxiv.org/abs/1507.02091}{{\ttfamily arXiv:1507.02091
  [hep-ph]}}.

\bibitem{Annala:2019puf}
E.~Annala, T.~Gorda, A.~Kurkela, J.~N\"attil\"a, and A.~Vuorinen, ``{Evidence
  for quark-matter cores in massive neutron stars}'',
  \href{http://dx.doi.org/10.1038/s41567-020-0914-9}{{\em Nature Phys.}
  {\bfseries 16} no.~9, (2020) },
  \href{http://arxiv.org/abs/1903.09121}{{\ttfamily arXiv:1903.09121
  [astro-ph.HE]}}.

\bibitem{ALargeIonColliderExperiment:2021puh}
{\bfseries ALICE} Collaboration, S.~Acharya {\em et~al.}, ``{Hypertriton
  Production in p-Pb Collisions at $\sqrt {s_{NN}}$=5.02\,\,TeV}'',
  \href{http://dx.doi.org/10.1103/PhysRevLett.128.252003}{{\em Phys. Rev.
  Lett.} {\bfseries 128} no.~25, (2022) },
  \href{http://arxiv.org/abs/2107.10627}{{\ttfamily arXiv:2107.10627
  [nucl-ex]}}.

\bibitem{Bellini:2019zqc}
F.~Bellini and A.~P. Kalweit, ``{Testing production scenarios for
  (anti-)(hyper-)nuclei with multiplicity-dependent measurements at the LHC}'',
  \href{http://dx.doi.org/10.5506/APhysPolB.50.991}{{\em Acta Phys. Polon. B}
  {\bfseries 50} (2019) }, \href{http://arxiv.org/abs/1907.06868}{{\ttfamily
  arXiv:1907.06868 [hep-ph]}}.

\bibitem{Haidenbauer:2019thx}
J.~Haidenbauer and I.~Vidana, ``{Structure of single-$\Lambda$ hypernuclei with
  chiral hyperon-nucleon potentials}'',
  \href{http://dx.doi.org/10.1140/epja/s10050-020-00055-6}{{\em Eur. Phys. J.
  A} {\bfseries 56} no.~2, (2020) },
  \href{http://arxiv.org/abs/1910.02695}{{\ttfamily arXiv:1910.02695
  [nucl-th]}}.

\bibitem{Fairbairn:2006gg}
M.~Fairbairn, A.~C. Kraan, D.~A. Milstead, T.~Sjostrand, P.~Z. Skands, and
  T.~Sloan, ``{Stable Massive Particles at Colliders}'',
  \href{http://dx.doi.org/10.1016/j.physrep.2006.10.002}{{\em Phys. Rept.}
  {\bfseries 438} (2007) },
  \href{http://arxiv.org/abs/hep-ph/0611040}{{\ttfamily arXiv:hep-ph/0611040}}.

\bibitem{Bruce:2018yzs}
R.~Bruce {\em et~al.}, ``{New physics searches with heavy-ion collisions at the
  CERN Large Hadron Collider}'',
  \href{http://dx.doi.org/10.1088/1361-6471/ab7ff7}{{\em J. Phys. G} {\bfseries
  47} no.~6, (2020) }, \href{http://arxiv.org/abs/1812.07688}{{\ttfamily
  arXiv:1812.07688 [hep-ph]}}.

\bibitem{CMSTracker:2005iuk}
{\bfseries CMS Tracker} Collaboration, W.~Adam {\em et~al.}, ``{The effect of
  highly ionising particles on the CMS silicon strip tracker}'',
  \href{http://dx.doi.org/10.1016/j.nima.2004.11.049}{{\em Nucl. Instrum. Meth.
  A} {\bfseries 543} no.~2-3, (2005) }.

\bibitem{Donato:1999gy}
F.~Donato, N.~Fornengo, and P.~Salati, ``{Anti-deuterons as a signature of
  supersymmetric dark matter}'',
  \href{http://dx.doi.org/10.1103/PhysRevD.62.043003}{{\em Phys. Rev. D}
  {\bfseries 62} (2000) },
  \href{http://arxiv.org/abs/hep-ph/9904481}{{\ttfamily arXiv:hep-ph/9904481}}.

\bibitem{Duperray:2005si}
R.~Duperray, B.~Baret, D.~Maurin, G.~Boudoul, A.~Barrau, L.~Derome,
  K.~Protasov, and M.~Buenerd, ``{Flux of light antimatter nuclei near Earth,
  induced by cosmic rays in the Galaxy and in the atmosphere}'',
  \href{http://dx.doi.org/10.1103/PhysRevD.71.083013}{{\em Phys. Rev. D}
  {\bfseries 71} (2005) },
  \href{http://arxiv.org/abs/astro-ph/0503544}{{\ttfamily
  arXiv:astro-ph/0503544}}.

\bibitem{Winkler:2020ltd}
M.~W. Winkler and T.~Linden, ``{Dark Matter Annihilation Can Produce a
  Detectable Antihelium Flux through $\bar{\Lambda}_b$ Decays}'',
  \href{http://dx.doi.org/10.1103/PhysRevLett.126.101101}{{\em Phys. Rev.
  Lett.} {\bfseries 126} no.~10, (2021) },
  \href{http://arxiv.org/abs/2006.16251}{{\ttfamily arXiv:2006.16251
  [hep-ph]}}.

\end{thebibliography}\endgroup
\newpage
\appendix
\section{The ALICE-USA Collaboration}
\label{app:collab}
\begin{flushleft}
N.~Alizadehvandchali$^{\rm 9}$,
N.~Apadula$^{\rm 5}$,
M.~Arslandok$^{\rm 13}$,
C.~Beattie$^{\rm 13}$,
R.~Bellwied$^{\rm 9}$,
J.T.~Blair$^{\rm 8}$,
F.~Bock$^{\rm 6}$,
H.~Bossi$^{\rm 13}$,
A.~Bylinkin$^{\rm 10}$,
H.~Caines$^{\rm 13}$,
I.~Chakaberia$^{\rm 5}$,
M.~Cherney$^{\rm 3}$,
T.M.~Cormier$^{\rm I,}$$^{\rm 6}$,
R.~Cruz-Torres$^{\rm 5}$,
P.~Dhankher$^{\rm 4}$,
D.U.~Dixit$^{\rm 4}$,
R.J.~Ehlers$^{\rm 5,6}$,
W.~Fan$^{\rm 5}$,
M.~Fasel$^{\rm 6}$,
F.~Flor$^{\rm 9}$,
A.N.~Flores$^{\rm 8}$,
D.R.~Gangadharan$^{\rm 9}$,
E.~Garcia-Solis$^{\rm 2}$,
A.~Gautam$^{\rm 10}$,
E.~Glimos$^{\rm 11}$,
V.~Gonzalez$^{\rm 12}$,
A.~Hamdi$^{\rm 5}$,
R.~Hannigan$^{\rm 8}$,
J.W.~Harris$^{\rm 13}$,
A.~Harton$^{\rm 2}$,
H.~Hassan$^{\rm 6}$,
L.B.~Havener$^{\rm 13}$,
C.~Hughes$^{\rm 11}$,
T.J.~Humanic$^{\rm 7}$,
A.~Hutson$^{\rm 9}$,
T.~Isidori$^{\rm 10}$,
B.~Jacak$^{\rm 4,5}$,
P.M.~Jacobs$^{\rm 5}$,
F.~Jonas$^{\rm 6}$,
A.~Khatun$^{\rm 10}$,
M.~Kim$^{\rm 4}$,
J.L.~Klay$^{\rm 1}$,
S.~Klein$^{\rm 5}$,
A.G.~Knospe$^{\rm 9}$,
Y.S.~Lai$^{\rm 5}$,
E.D.~Lesser$^{\rm 4}$,
I.~Likmeta$^{\rm 9}$,
A.~Liu$^{\rm 4}$,
C.~Loizides$^{\rm 6}$,
C.~Markert$^{\rm 8}$,
J.L.~Martinez$^{\rm 9}$,
A.S.~Menon$^{\rm 9}$,
J.D.~Mulligan$^{\rm 5}$,
A.I.~Nambrath$^{\rm 4}$,
C.~Nattrass$^{\rm 11}$,
N.~Novitzky$^{\rm 6}$,
A.C.~Oliveira Da Silva$^{\rm 11}$,
M.H.~Oliver$^{\rm 13}$,
L.~Pinsky$^{\rm 9}$,
M.~P\l osko\'{n}$^{\rm 5}$,
M.G.~Poghosyan$^{\rm 6}$,
C.A.~Pruneau$^{\rm 12}$,
R.E.~Quishpe$^{\rm 9}$,
S.~Ragoni$^{\rm 3}$,
K.F.~Read$^{\rm 6,11}$,
O.V.~Rueda$^{\rm 9}$,
D.~Sarkar$^{\rm 12}$,
M.H.P.~Sas$^{\rm 13}$,
J.~Schambach$^{\rm 6}$,
N.V.~Schmidt$^{\rm 6}$,
A.R.~Schmier$^{\rm 11}$,
J.E.~Seger$^{\rm 3}$,
O.~Sheibani$^{\rm 9}$,
N.~Smirnov$^{\rm 13}$,
J.~Song$^{\rm 9}$,
P.J.~Steffanic$^{\rm 11}$,
J.D.~Tapia Takaki$^{\rm 10}$,
C.~Terrevoli$^{\rm 9}$,
D.~Thomas$^{\rm 8}$,
A.R.~Timmins$^{\rm 9}$,
S.A.~Voloshin$^{\rm 12}$,
S.L.~Weyhmiller$^{\rm 13}$,
J.R.~Wright$^{\rm 8}$

\section*{Affiliation Notes} 
$^{\rm I}$ Deceased\\

\section*{Collaboration Institutes}
$^{1}$ California Polytechnic State University, San Luis Obispo, California\\
$^{2}$ Chicago State University, Chicago, Illinois\\
$^{3}$ Creighton University, Omaha, Nebraska\\
$^{4}$ Department of Physics, University of California, Berkeley, California\\
$^{5}$ Lawrence Berkeley National Laboratory, Berkeley, California\\
$^{6}$ Oak Ridge National Laboratory, Oak Ridge, Tennessee\\
$^{7}$ Ohio State University, Columbus, Ohio\\
$^{8}$ The University of Texas at Austin, Austin, Texas\\
$^{9}$ University of Houston, Houston, Texas\\
$^{10}$ The University of Kansas, Lawrence, Kansas\\
$^{11}$ University of Tennessee, Knoxville\\
$^{12}$ Wayne State University, Detroit, Michigan\\
$^{13}$ Yale University, New Haven, Connecticut\\
 
\end{flushleft}  

\end{document}